\documentclass[aps,prd,superscriptaddress,preprint,tightenlines,nofootinbib]{revtex4}

\usepackage{graphicx} 
\usepackage{dcolumn}  
\usepackage{bm}       

\begin{document}

\newcommand{\DECAY}{$D_s^+\to K^-\pi^+\pi^+$}
\newcommand{\PM}{$\pm$}
\newcommand{\FIGDIR}{.}

\preprint{CLNS 09/2048}       
\preprint{CLEO 09-01}         


\title{Dalitz Plot Analysis of $D_s^+\to K^+K^-\pi^+$}

\author{R.~E.~Mitchell}
\author{M.~R.~Shepherd}
\affiliation{Indiana University, Bloomington, Indiana 47405, USA }
\author{D.~Besson}
\affiliation{University of Kansas, Lawrence, Kansas 66045, USA}
\author{T.~K.~Pedlar}
\author{J.~Xavier}
\affiliation{Luther College, Decorah, Iowa 52101, USA}
\author{D.~Cronin-Hennessy}
\author{K.~Y.~Gao}
\author{J.~Hietala}
\author{Y.~Kubota}
\author{T.~Klein}
\author{R.~Poling}
\author{A.~W.~Scott}
\author{P.~Zweber}
\affiliation{University of Minnesota, Minneapolis, Minnesota 55455, USA}
\author{S.~Dobbs}
\author{Z.~Metreveli}
\author{K.~K.~Seth}
\author{B.~J.~Y.~Tan}
\author{A.~Tomaradze}
\affiliation{Northwestern University, Evanston, Illinois 60208, USA}
\author{J.~Libby}
\author{L.~Martin}
\author{A.~Powell}
\author{C.~Thomas}
\author{G.~Wilkinson}
\affiliation{University of Oxford, Oxford OX1 3RH, UK}
\author{H.~Mendez}
\affiliation{University of Puerto Rico, Mayaguez, Puerto Rico 00681}
\author{J.~Y.~Ge}
\author{D.~H.~Miller}
\author{I.~P.~J.~Shipsey}
\author{B.~Xin}
\affiliation{Purdue University, West Lafayette, Indiana 47907, USA}
\author{G.~S.~Adams}
\author{D.~Hu}
\author{B.~Moziak}
\author{J.~Napolitano}
\affiliation{Rensselaer Polytechnic Institute, Troy, New York 12180, USA}
\author{K.~M.~Ecklund}
\affiliation{Rice University, Houston, TX 77005, USA}
\author{Q.~He}
\author{J.~Insler}
\author{H.~Muramatsu}
\author{C.~S.~Park}
\author{E.~H.~Thorndike}
\author{F.~Yang}
\affiliation{University of Rochester, Rochester, New York 14627, USA}
\author{M.~Artuso}
\author{S.~Blusk}
\author{S.~Khalil}
\author{R.~Mountain}
\author{K.~Randrianarivony}
\author{N.~Sultana}
\author{T.~Skwarnicki}
\author{S.~Stone}
\author{J.~C.~Wang}
\author{L.~M.~Zhang}
\affiliation{Syracuse University, Syracuse, New York 13244, USA}
\author{G.~Bonvicini}
\author{D.~Cinabro}
\author{M.~Dubrovin}
\author{A.~Lincoln}
\author{M.~J.~Smith}
\affiliation{Wayne State University, Detroit, Michigan 48202, USA}
\author{P.~Naik}
\author{J.~Rademacker}
\affiliation{University of Bristol, Bristol BS8 1TL, UK}
\author{D.~M.~Asner}
\author{K.~W.~Edwards}
\author{J.~Reed}
\author{A.~N.~Robichaud}
\author{G.~Tatishvili}
\author{E.~J.~White}
\affiliation{Carleton University, Ottawa, Ontario, Canada K1S 5B6}
\author{R.~A.~Briere}
\author{H.~Vogel}
\affiliation{Carnegie Mellon University, Pittsburgh, Pennsylvania 15213, USA}
\author{P.~U.~E.~Onyisi}
\author{J.~L.~Rosner}
\affiliation{Enrico Fermi Institute, University of
Chicago, Chicago, Illinois 60637, USA}
\author{J.~P.~Alexander}
\author{D.~G.~Cassel}
\author{J.~E.~Duboscq}\thanks{Deceased}
\author{R.~Ehrlich}
\author{L.~Fields}
\author{L.~Gibbons}
\author{R.~Gray}
\author{S.~W.~Gray}
\author{D.~L.~Hartill}
\author{B.~K.~Heltsley}
\author{D.~Hertz}
\author{J.~M.~Hunt}
\author{J.~Kandaswamy}
\author{D.~L.~Kreinick}
\author{V.~E.~Kuznetsov}
\author{J.~Ledoux}
\author{H.~Mahlke-Kr\"uger}
\author{J.~R.~Patterson}
\author{D.~Peterson}
\author{D.~Riley}
\author{A.~Ryd}
\author{A.~J.~Sadoff}
\author{X.~Shi}
\author{S.~Stroiney}
\author{W.~M.~Sun}
\author{T.~Wilksen}
\affiliation{Cornell University, Ithaca, New York 14853, USA}
\author{J.~Yelton}
\affiliation{University of Florida, Gainesville, Florida 32611, USA}
\author{P.~Rubin}
\affiliation{George Mason University, Fairfax, Virginia 22030, USA}
\author{N.~Lowrey}
\author{S.~Mehrabyan}
\author{M.~Selen}
\author{J.~Wiss}
\affiliation{University of Illinois, Urbana-Champaign, Illinois 61801, USA}
\collaboration{CLEO Collaboration}
\noaffiliation

\date{6 March 2009}

\begin{abstract}
We perform a Dalitz plot analysis of the decay $D_s^+\to K^+K^-\pi^+$
with the CLEO-c data set of 586~pb$^{-1}$ of $e^+e^-$ collisions
accumulated at $\sqrt{s} = 4.17$~GeV.  This corresponds to
about 0.57 million $D_s^\pm D_s^{*\mp}$ pairs from which we select 14400
candidates with a background of roughly 15\%.
In contrast to previous measurements we find
good agreement with our data only by including an
additional $f_0(1370)\pi^+$ contribution.
We measure the magnitude, phase, and fit fraction of
$K^*(892)^0 K^+$, 
$\phi(1020)\pi^+$, 
$K_0^*(1430)K^+$, 
$f_0(980)\pi^+$, 
$f_0(1710)\pi^+$,
and 
$f_0(1370)\pi^+$ contributions and limit the
possible contributions of other $KK$ and $K\pi$ resonances
that could appear in this decay.
\end{abstract}


\maketitle


\section{Introduction}
\label{sec:introduction}

The decay $D_s^+ \to K^+K^-\pi^+$
is among the largest known  
branching fractions for the $D_s$ meson.
For some time the mode $D_s^+ \to \phi(1020) \pi^+$ was used as the normalizing mode
for $D_s$ decay branching fractions, 
typically done by choosing events with the $K^+K^-$ invariant mass
near the narrow $\phi(1020)$ peak.
Observation of a large contribution from $D_s^+ \to f_0(980) \pi^+$ \cite{E687}
makes the selection of $D_s^+ \to \phi(1020) \pi^+$ dependent
on the range of $K^+K^-$ invariant mass chosen; the observed
yield of non-$\phi$ contributions can be larger than 10\%~\cite{cleo-Dshad}.
This is an unacceptably large uncertainty for a normalizing mode and
we proposed~\cite{cleo-Dshad} that the branching fraction for 
$D_s^+ \to K^+K^-\pi^+$ in the neighborhood of the $\phi$ peak, without any attempt to  
identify the $\phi \pi^+$ component as such, could be used for $D_s$ normalization.   
Relating  the $D_s^+ \to K^+K^-\pi^+$ branching fraction in~\cite{cleo-Dshad} to the rates
for such phase space-restricted subsets requires an understanding of the 
resonance contributions to the final state.
The only published Dalitz plot analysis~\cite{Dalitz} has been done by E687 \cite{E687} 
using 701 signal events.
The FOCUS Collaboration has studied this decay in a Dalitz plot analysis
in an unpublished thesis \cite{FOCUS_THESIS}
and a conference presentation \cite{FOCUS}.

	Here we describe a Dalitz plot analysis of $D_s^+ \to K^+K^-\pi^+$
using the CLEO-c data set which yields a sample of over $12,000$ signal candidates.
Charge conjugation is implied throughout except where explicitly mentioned.
The next section describes our experimental techniques,
the third section gives our Dalitz plot analysis formalism, the fourth
describes our fits to the data, and there is a brief conclusion.


\section{Experimental technique}
\label{sec:detector}

CLEO-c is a general purpose detector which includes a tracking system
for measuring momenta and specific ionization of charged particles,
a Ring Imaging Cherenkov detector to aid particle identification,
and a CsI calorimeter for detection of electromagnetic showers.
These components are immersed in a magnetic field of 1~T,
provided by a superconducting solenoid, and surrounded by a muon detector.
The CLEO-c detector is described in detail elsewhere~\cite{CLEO-c}.

We reconstruct the $D_s^+ \to K^+K^-\pi^+$ decay
using three tracks measured in the tracking system.
Charged tracks satisfy standard goodness of fit quality requirements~\cite{HadronicBF}.
Pion and kaon candidates are required to have specific ionization,
$dE/dx$, in the main drift chamber within four standard
deviations of the expected value at the measured momentum.

We use two kinematic variables to select $D_s^+ \to K^+K^-\pi^+$  decays,
the candidate invariant mass
\begin{equation}
      m_{\rm inv} \equiv m(K^+K^-\pi^+)~~~{\rm or }~~~\Delta m_{\rm inv} = m_{\rm inv} - m_{D_s},
\end{equation}
and the beam constrained mass
\begin{equation}
   m_{\rm BC} = \sqrt{E^2_{\rm beam}-p^2_D}~~~{\rm or }~~~\Delta m_{\rm BC} = m_{\rm BC}-m_{\rm BC}(D_s),
\end{equation}
where 
$m_{D_s}$=1968.2~MeV/$c^2$~\cite{PDG-2006} is the $D_s$ mass,
$E_{\rm beam}$ is the beam energy,
$p_D$ is the momentum of reconstructed $D_s^+$ candidate, and
$m_{\rm BC}(D_s)$ = 2040.25~MeV/$c^2$ is the expected $m_{\rm BC}$ value of the $D_s$ meson 
in the process $e^+e^- \to D_s^*D_s$ at $\sqrt{s} = 4170$~MeV. 
We require
$|\Delta m_{\rm inv}| < 2 \sigma(m_{\rm inv})$,
$|\Delta m_{\rm BC}|  < 2 \sigma(m_{\rm BC})$, where the resolutions
$\sigma(m_{\rm inv}) = 4.8$~MeV/$c^2$ (4.79$\pm$0.05~MeV/$c^2$ in single Gaussian fit), and
$\sigma(m_{\rm BC})  = 2$~MeV/$c^2$   (1.89$\pm$0.02~MeV/$c^2$) represent the widths of
the signal peak in the two dimensional distribution.
When there are multiple $D_s$-meson candidates in a single event
we select the one with smallest $\Delta m_{\rm BC}$ value.

We use a kinematic fit to all 3-track combinations which enforces a
common vertex and $D_s^+$ mass constraint.
The kinematic fit-corrected 4-momenta of all 3 particles are used to
calculate invariant masses for further Dalitz plot analysis.  The resolution
on the resonance invariant mass is almost always better than 5~MeV/$c^2$.

\begin{figure}[th]
  \includegraphics[width=160mm]{\FIGDIR/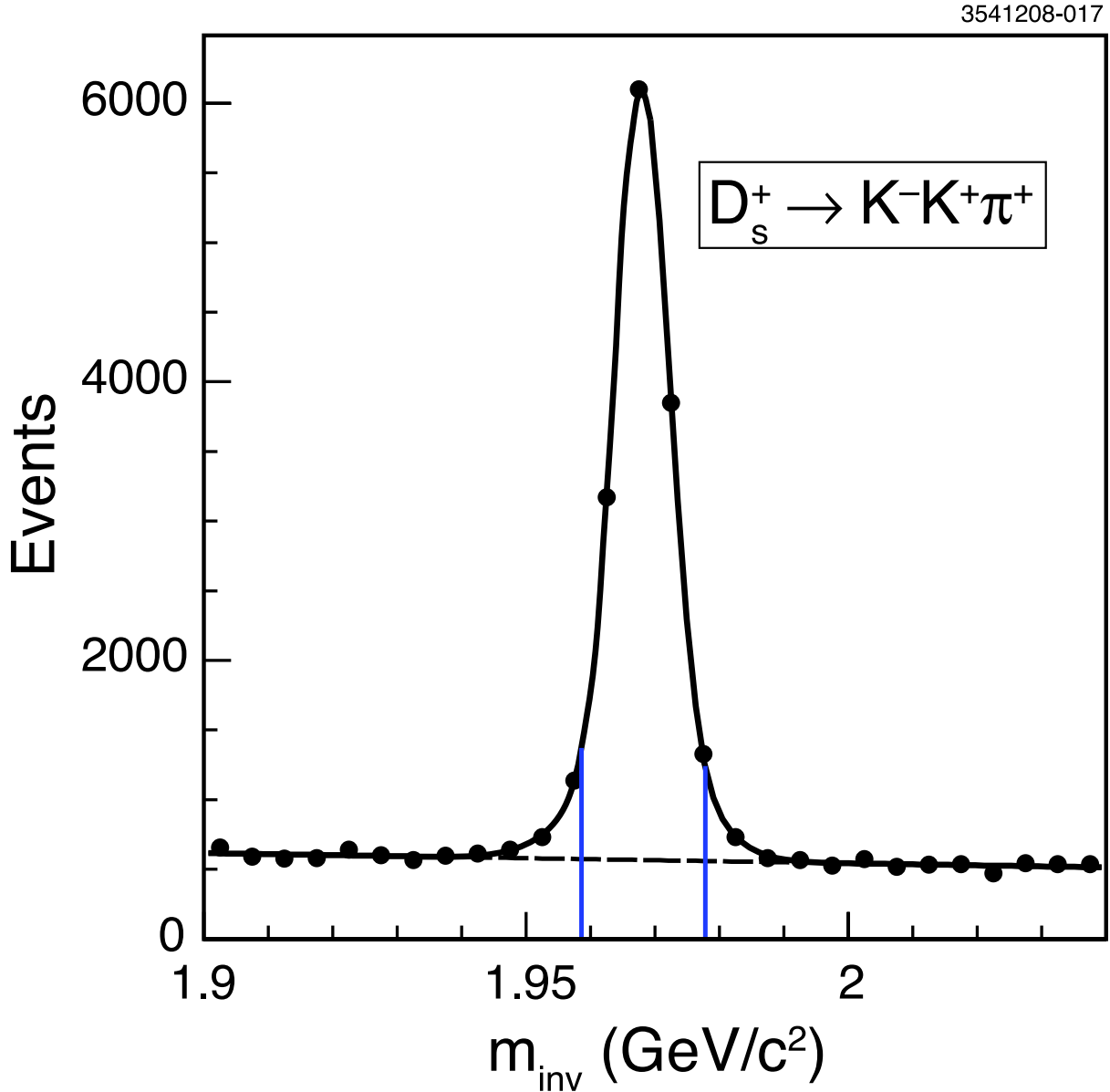}
  \caption{\label{fig:selection_2} The $m_{inv}$ distribution.
                                   The vertical (online blue) lines show the $\pm 2\sigma$ signal region.} 
\end{figure}

After all requirements,
we select 14400 candidate events for the Dalitz plot analysis.
The fraction of background, 15.1\%, in this sample is estimated from
the fits to the $m_{\rm inv}$ spectrum shown in Figure~\ref{fig:selection_2}.
In most fits to the Dalitz plot
we constrain the value of the signal fraction, $f_{\rm sig}=84.90\pm0.15$\%.
In cross checks we use a set of sub-samples,
splitting the data  by time of observation and by sign
of $D_s$-meson charge, $D_s^+$ and $D_s^-$.
We also consider samples with tight 
($1 \times 1$ standard deviations in $m_{\rm BC}$ and $m_{\rm inv}$)
and loose ($3 \times 3$ three standard deviations) selection
versus the standard selection,
as well as samples of $D_s$ mesons produced in $D_s^* \to D_s \gamma$ decays, selected 
with a displaced signal box using 
$m_{\rm BC}$ low  band ($|m_{\rm BC}-2025~{\rm MeV/c^2}| < 4\sigma(m_{\rm BC})$) and
$m_{\rm BC}$ high band ($|m_{\rm BC}-2060~{\rm MeV/c^2}| < 4\sigma(m_{\rm BC})$).

To determine the efficiency we use a signal Monte-Carlo (MC)
\cite{EVTGEN} simulation 
where one of the charged $D_s$ mesons decays in the $KK\pi$ mode
uniformly in the phase space, while the other $D_s$ meson decays
in all known modes with relevant branching fractions.
In total we generated $10^6$ $D_s^+$ and $D_s^-$ signal decays.
These underlying events are input to the CLEO-c detector
simulation and processed with the CLEO-c reconstruction package.
The MC-generated events are required to pass the same selection requirements as data
selected in the signal box.
We only select the signal-side $D_s$ mesons which decay uniformly in the phase space,
separating them by charge.

\begin{figure}[th]
  \includegraphics[width=160mm]{\FIGDIR/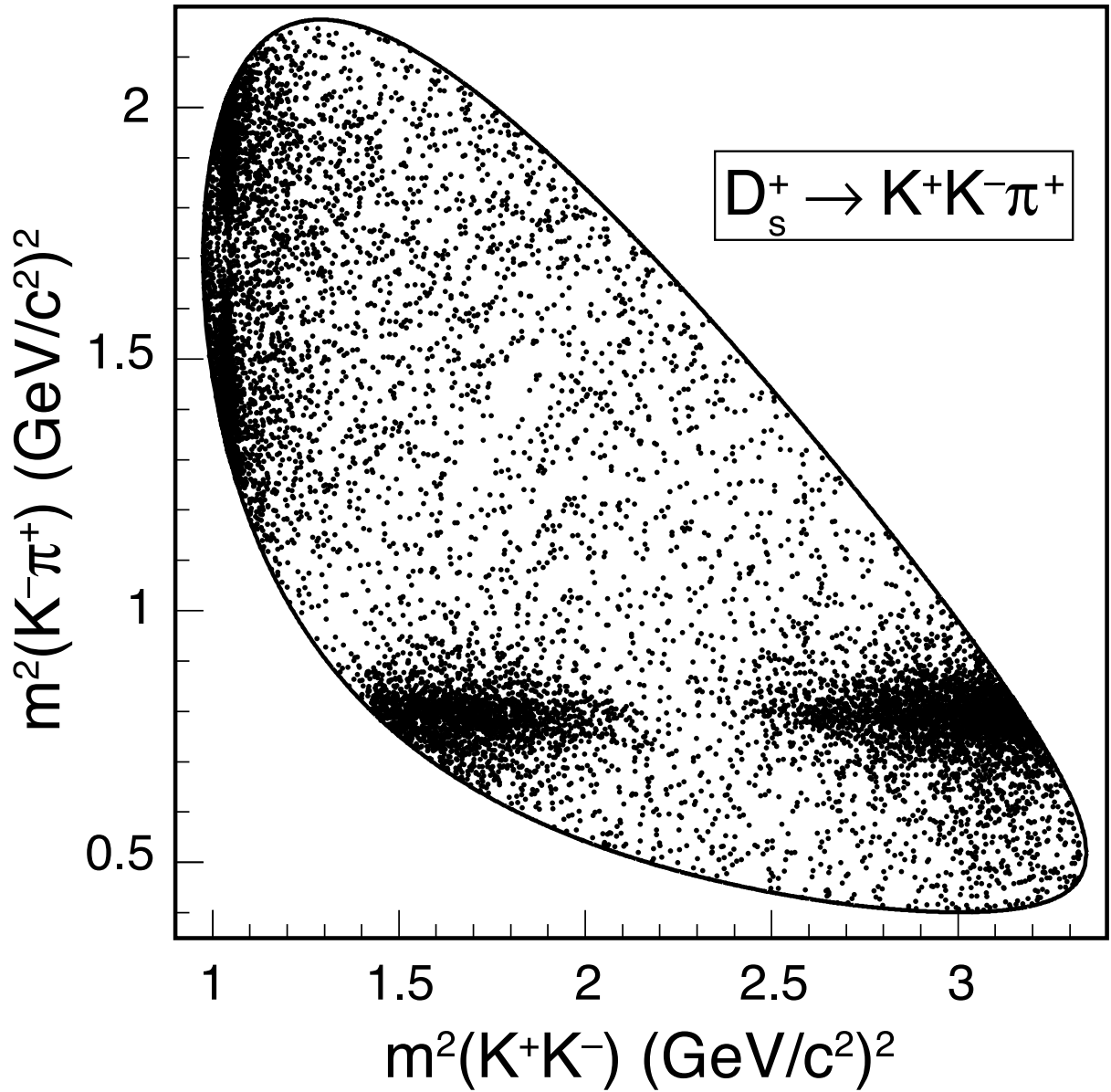}       \hfill
  \caption{\label{fig:DP_data} The Dalitz plot for the data.}
\end{figure}

We analyze events on the Dalitz plot by choosing
$x = m^2(K^+K^-)$ and
$y = m^2(K^-\pi^+)$ as the independent ($x,y$) variables.
The third variable $z = m^2(K^+\pi^+)$ is dependent on $x$ and $y$
through energy and momentum conservation.
We do not expect any resonant sub-structure in the $K^+\pi^+$ invariant mass;
with these Dalitz plot variables any structure in $z$ is due to reflections
of structures in $x$ and $y$.
Figure~\ref{fig:DP_data} shows the Dalitz plot.
Besides the clear $\phi(1020)$ and $K^*(892)$ signal, 
no other narrow features are clearly observed. 
The variation of the population density along the resonance 
band clearly indicates that these resonances are spin one
as the amplitude for a spin-one resonance should have a node
in the middle of its band.
There is a significant population density in the node region of 
the $\phi(1020)$ resonance, indicating that there is likely
to be an additional contribution.

To parametrize the efficiency, $\varepsilon (x,y)$,
we use a third-order polynomial function 
with respect to the arbitrary 
point ($x_c$, $y_c$)=(2,1)~(GeV/$c^2$)$^2$ on the Dalitz plot times
threshold functions in each of the Dalitz variables to account for
the loss of efficiency at the edges of the Dalitz plot, such that
\begin{equation}
\label{eqn:efficiency}
      \varepsilon (x,y) = \varepsilon_{poly}(x,y) T(x) T(y) T(z(x,y)).
\end{equation}
With
$\hat{x} = x-x_c$ and
$\hat{y} = y-y_c$,
the efficiency is the product of the polynomial function,
\begin{equation}
\label{eqn:efficiency_poly}
\varepsilon_{poly}(x,y) = 1 + E_x\hat{x}+E_y\hat{y} 
                      + E_{x^2}\hat{x}^2+E_{y^2}\hat{y}^2
                      + E_{x^3}\hat{x}^3+E_{y^3}\hat{y}^3
                      + E_{xy}\hat{x}\hat{y} + 
                        E_{x^2y}\hat{x}^2\hat{y}+E_{xy^2}\hat{x}\hat{y}^2,
\end{equation}
For each Dalitz plot variable, $v$  ($\equiv x, y$ or $z$)
the threshold function is sine-like with
\begin{equation}
\label{eqn:threshold_factor}
   T(v) = \left\{
\begin{array}{ll}[E_{c,v} + (1-E_{c,v})] \times
                 \sin( E_{{\rm th},v}\times|v-v_{max}| ), 
                                  & {\rm ~~~at~~ } 0<E_{{\rm th},v}\times|v-v_{\rm max}|<\pi/2, \\
                 1              , & {\rm ~~~at~~ } E_{{\rm th},v}\times|v-v_{\rm max}| \geq \pi/2, \\
\end{array} \right.
\end{equation}
All polynomial coefficients, 
$E_x$, $E_y$, $E_{x^2}$, $E_{y^2}$, $E_{x^3}$, $E_{y^3}$, 
$E_{xy}$, $E_{x^2y}$, $E_{xy^2}$, $E_{c,v}$, and $E_{th,v}$ 
are fit parameters.
Each variable $v$ has two thresholds, $v_{\rm min}$ and $v_{\rm max}$.
We expect low efficiency in the regions $v \approx v_{\rm max}$ only, where
one of three particles is produced with zero momentum in the $D_s$ meson rest frame
and thus has a small momentum in the laboratory frame.

The simulated signal sample
is used to determine the efficiency.
Table~\ref{tab:Efficiency}
\begin{table}[th]
\caption{\label{tab:Efficiency} Fit parameters for describing the efficiency across
                                the Dalitz plot.}
\begin{center}
\begin{tabular}{lc}
\hline
\hline
Parameter       & Value \\ \hline
$E_x$       &  0.023\PM0.012 \\
$E_y$       &  0.037\PM0.014 \\
$E_{x2}$    &--0.307\PM0.014 \\
$E_{xy}$    &--0.526\PM0.034 \\
$E_{y2}$    &--0.201\PM0.034 \\
$E_{x3}$    &  0.262\PM0.026 \\
$E_{x2y}$   &  0.953\PM0.078 \\
$E_{xy2}$   &  0.887\PM0.098 \\
$E_{y3}$    &  0.004\PM0.051 \\
\hline
$E_{th,x}$  &  3.23\PM0.18 \\
$E_{th,y}$  &  2.53\PM0.13 \\
$E_{th,z}$  &  2.61\PM0.13 \\
$E_{c,x}$   &  0.166\PM0.042 \\
$E_{c,y}$   &  0.320\PM0.034 \\
$E_{c,z}$   &  0.338\PM0.034 \\
\hline				  	  	   
\hline
\end{tabular}
\end{center}
\end{table}
shows the results of the fit to the entire signal MC sample
of $D_s^+ \to K^+K^-\pi^+$  events selected on the Dalitz plot.
The polynomial function with threshold factors describes the efficiency
shape very well for our sample. 
We also fit separately the signal MC sub-samples for
$D_s^+ \to K^+K^-\pi^+$ and $D_s^- \to K^-K^+\pi^-$ decays,
for simulations of early and late datasets, and for tight and loose signal boxes.  
In cross-checks with sub-samples we fix the threshold parameters to their values 
from the central fit in order to remove correlations with other polynomial parameters.
We find that the variation of the efficiency polynomial parameters 
is small compared to their statistical uncertainties.
In fits to data we use this efficiency shape with fixed parameters,
and constrained variation is allowed as a systematic check.

\begin{figure}[th]
  \includegraphics[width=72mm]{\FIGDIR/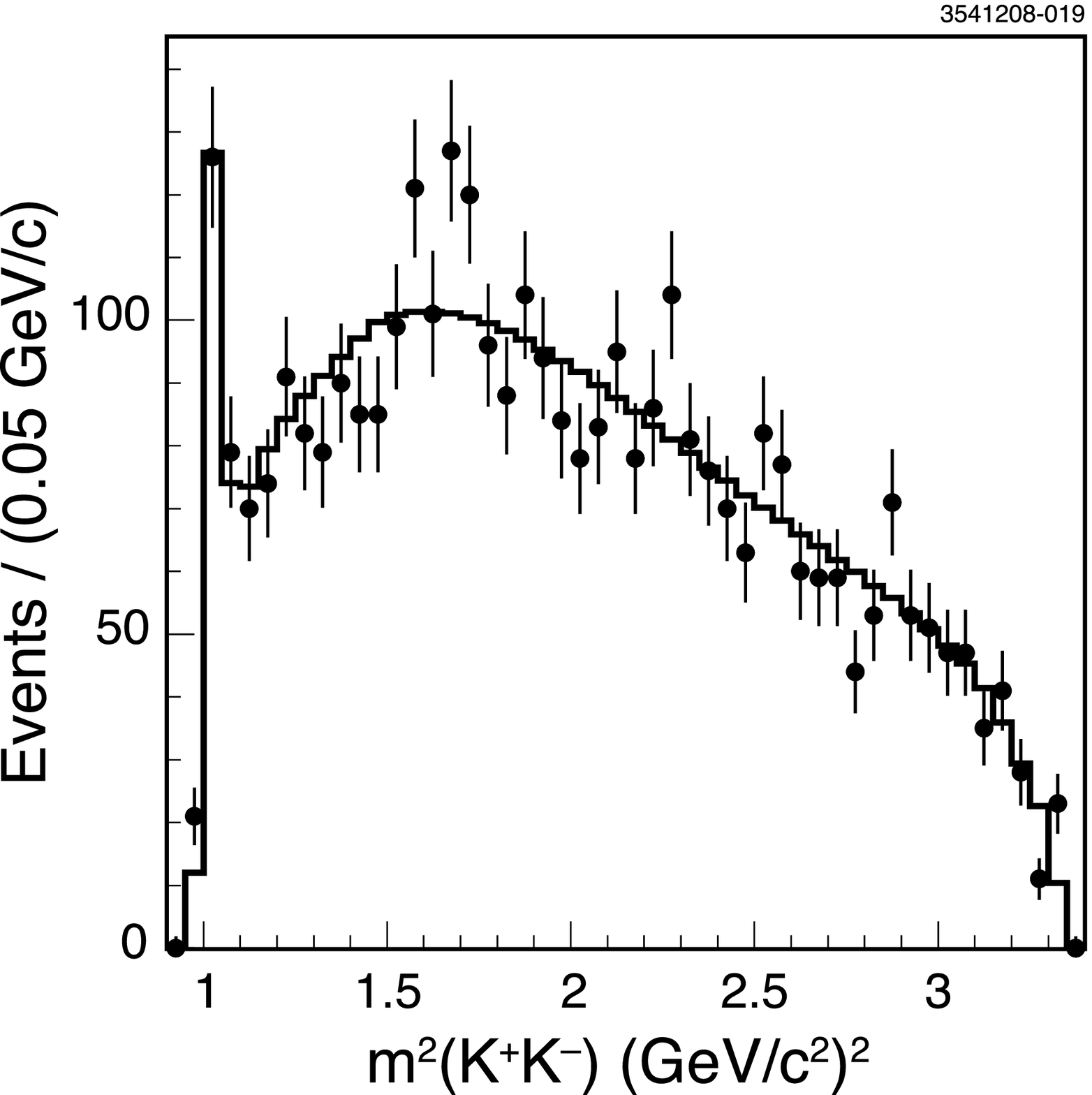} \hfill
  \includegraphics[width=72mm]{\FIGDIR/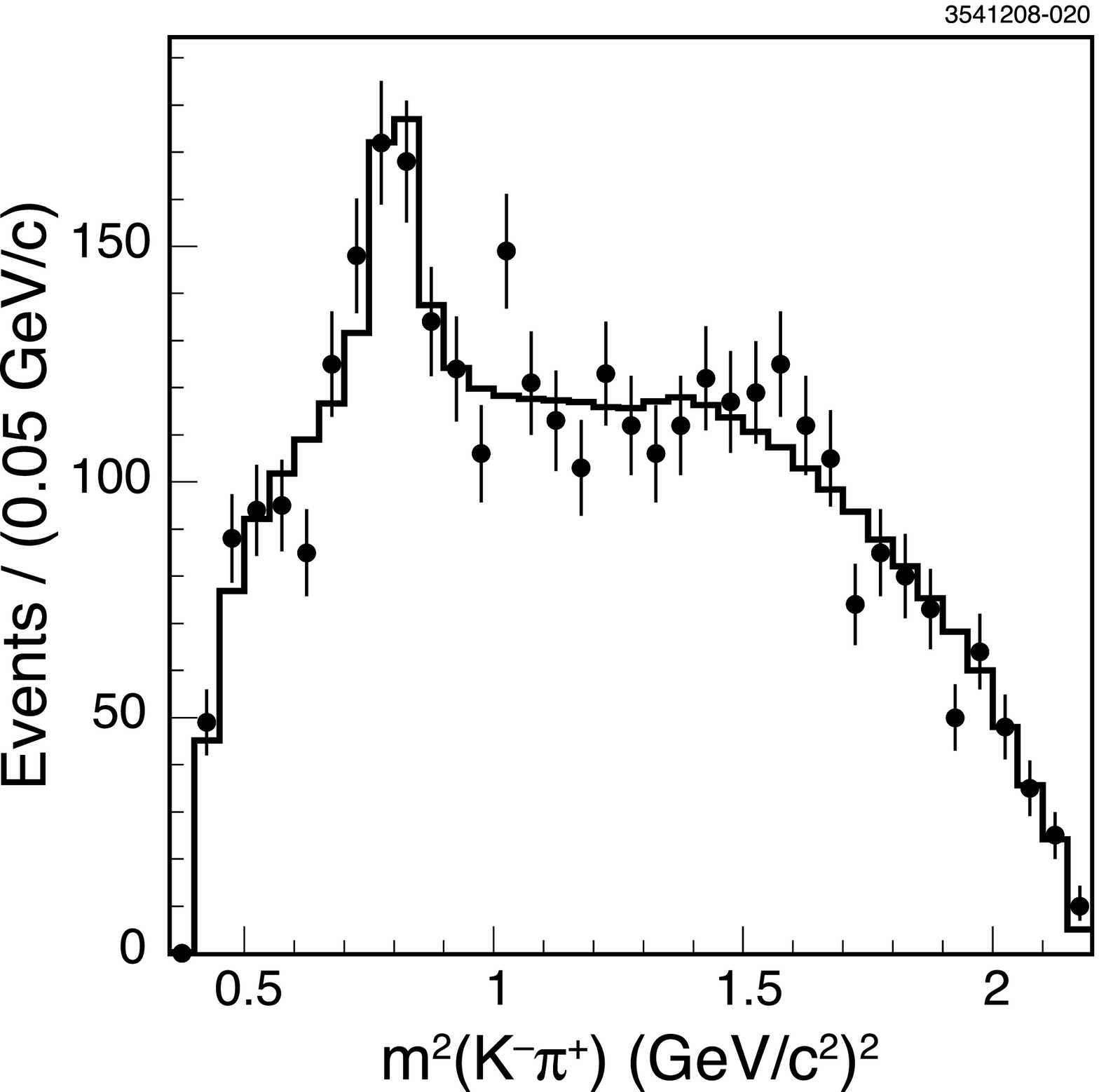} \hfill
  \includegraphics[width=72mm]{\FIGDIR/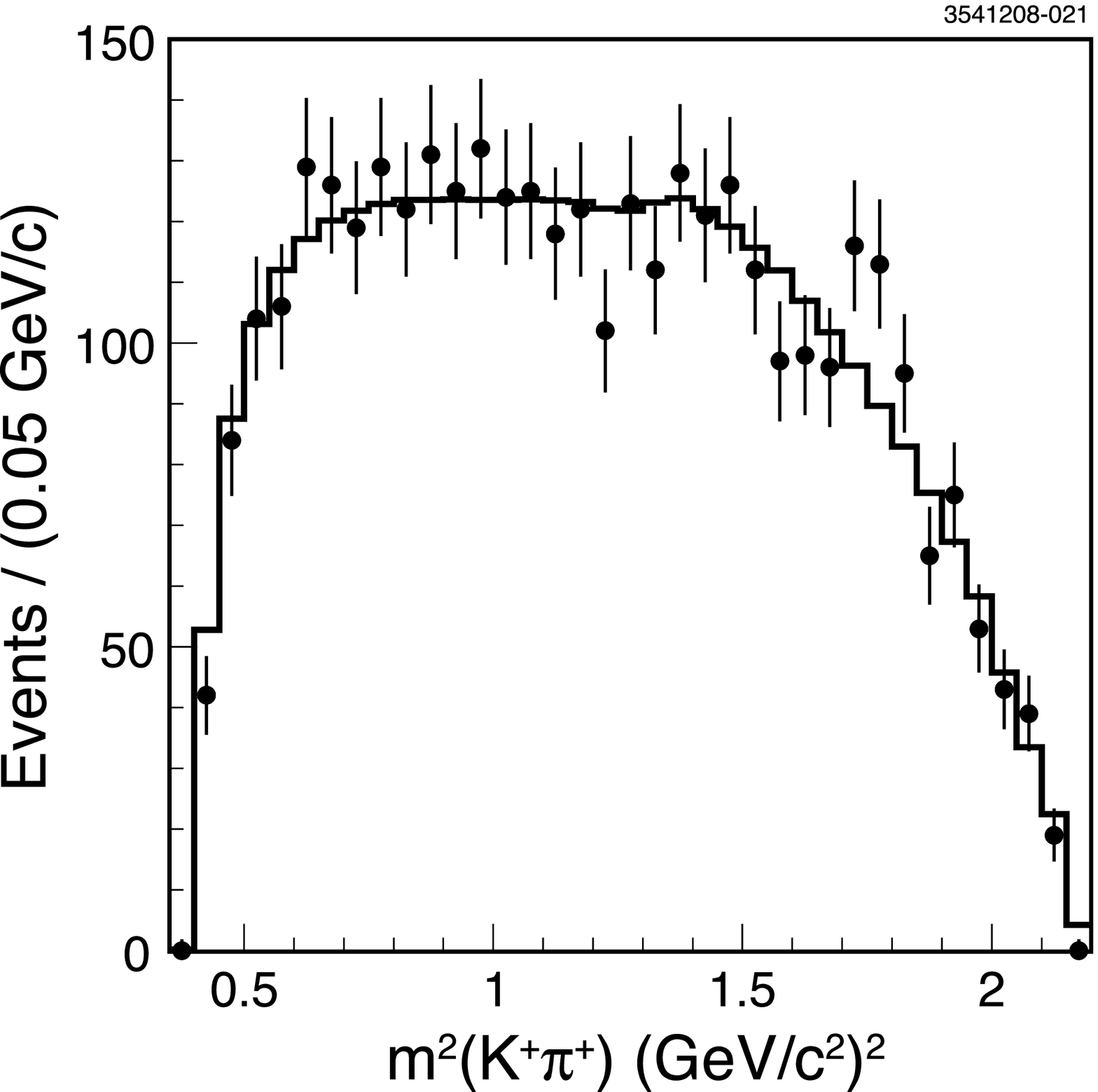}
  \caption{\label{fig:background_proj} Projections of the fit to the background shape
                                       described in the text, line,
                                       displayed over the data, dots, in the background box.} 
\end{figure}

The shape for the background on the Dalitz plot is estimated
using data events from a $m_{\rm BC}$ sideband region,
$|m_{\rm BC}-1900 {\rm MeV/c}^2| < 5\sigma(m_{\rm BC})$.
We only consider events from the low mass $m_{\rm BC}$ sideband
as the high mass sideband is contaminated by signal
events due to initial state radiation.
To parametrize the background shape on the Dalitz plot
we employ a function similar to that used for the efficiency, shown in 
Eq.~\ref{eqn:efficiency_poly}.
We add incoherently to the polynomial
two peaking contributions to represent $K^*(892)$ and $\phi(1020)$
contributions described with Breit-Wigner functions
with floating normalization coefficients, $B_{K^*}$ and $B_{\phi}$, respectively.
Figure~\ref{fig:background_proj} and Table~\ref{tab:Background} 
\begin{table}[th]
\caption{\label{tab:Background} Fit parameters for the background sample.
                                Values in parentheses show an uncertainty or 
                                variation of the last significant digits.}
\begin{center}
\begin{tabular}{lc}
\hline
\hline
Parameter       & Value \\ \hline
  $B_x$         &--0.23\PM0.11 \\
  $B_y$         &  0.06\PM0.13 \\
  $B_{x2}$      &--0.29\PM0.12 \\
  $B_{xy}$      &--0.99\PM0.29 \\
  $B_{y2}$      &--0.47\PM0.32 \\
  $B_{x3}$      &  0.77\PM0.23 \\
  $B_{x2y}$     &  1.98\PM0.67 \\
  $B_{xy2}$     &  2.24\PM0.84 \\
  $B_{y3}$      &  0.56\PM0.47 \\
  $B_{\phi}$    &  0.000161(23)    \\
  $B_{K^*}$     &  0.00144(28)     \\
\hline						    
\hline
\end{tabular}
\end{center}
\end{table}
show results of the fit to the background polynomial function 
for our sample.  We also consider
the variation of the background shape parameters for sub-samples,
split for  $D_s^+$ and $D_s^-$, for earlier and later datasets, and
for tight and loose cuts on background selection box. 
The variation of the shape parameters is small 
compared to their statistical uncertainties.  
Furthermore, in fits to data
we use the background shape with fixed parameters,
and constrained variation is allowed as a systematic cross check.
We also allow the size of the narrow resonance contributions to the
background to float freely as a systematic variation.

\section{Formalism}
\label{sec:formalism}

This Dalitz plot analysis employs the
techniques and formalism described in Ref.~\cite{Tim}
that have been applied in many other CLEO analyses.
We use an unbinned maximum likelihood fit that minimizes
the sum over $N$ events:
\begin{equation}
\label{eqn:LogL}
        \mathcal{L} = -2\sum_{n=1}^{N} \log {\cal P}(x_n,y_n),
\end{equation}
where ${\cal P}(x,y)$ is the probability density function (p.d.f.), 
which depends on
the event sample being fit,
\begin{equation}
\label{eqn:PDF}
{\cal P}(x,y) = \left\{
           \begin{array}{ll}
                  {\mathcal N}_{\varepsilon}\varepsilon(x,y) & {\rm for~efficiency;} \\
                  {\mathcal N}_B B(x,y)           & {\rm for~background;} \\
                  f_{\rm{sig}} {\mathcal N}_S |{\mathcal M}(x,y)|^2 \varepsilon(x,y)
                + (1-f_{\rm{sig}}) {\mathcal N}_B B(x,y)
                                   & {\rm for~signal.}
            \end{array}
              \right.
\end{equation}
The shapes for the efficiency, $\varepsilon(x,y)$, 
and background, $B(x,y)$, are discussed in the previous section.
The signal p.d.f. is proportional to the
efficiency-corrected matrix element squared, $|{\mathcal M}(x,y)|^2$.
As described above, the signal fraction, $f_{\rm{sig}}$, is defined from the invariant mass spectrum.
The background term has a relative $(1-f_{\rm{sig}})$ fraction.
The efficiency, signal, and background fractions are normalized separately,
$1/{\mathcal N}_{\varepsilon} = \int \varepsilon(x,y) dx dy$,
$1/{\mathcal N}_S = \int |{\mathcal M}(x,y)|^2 \varepsilon(x,y) dx dy$,
$1/{\mathcal N}_B = \int B(x,y) dx dy$, which provides the overall p.d.f. normalization,
$\int {\cal P}(x,y) dx dy =1$.
The matrix element is a sum of partial amplitudes,
\begin{equation}
\label{eqn:MatrixElement}
{\mathcal M} = \sum_R c_R  \times {\cal W}_R  \times \Omega_R  \times {\cal F}_D^L  \times {\cal F}_R^L,
\end{equation}
where ${\cal W}_R$ depends on the spin of resonance $R$.
The factor $\Omega_R$ is the angular distribution for the resonance, and the factors 
${\cal F}_D^L$ and ${\cal F}_R^L$ are the Blatt-Weisskopf angular momentum barrier-penetration 
factors \cite{Blatt-Weisskopf}.
In our standard fit the complex factor $c_R = a_R e^{i\phi_R}$
is represented by two real numbers, an amplitude $a_R$
and a phase $\phi_R$. These are included in the list of fit parameters
and can be left to float freely or fixed.

Assuming the decay chain $d\to Rc \to abc$ we may write the angular distribution,
\begin{eqnarray}
   \label{eqn:angular_distributions}
    \Omega_R^{L=0}&=& 1, \nonumber \\
    \Omega_R^{L=1}&=& m^2_{bc} - m^2_{ac} + \frac{(m^2_d - m^2_c)(m^2_a - m^2_b)}{m^2_{ab}}, \\
    \Omega_R^{L=2}&=&[\Omega_R^{L=1}]^2 - \frac{1}{3}
                 \bigg( m^2_{ab} - 2 m^2_d - 2 m^2_c + \frac{(m^2_d - m^2_c)^2}{m^2_{ab}} \bigg)    
                 \bigg( m^2_{ab} - 2 m^2_a - 2 m^2_b + \frac{(m^2_a - m^2_b)^2}{m^2_{ab}} \bigg), 
                          \nonumber 
\end{eqnarray}
where $m_d$ is the mass of the decaying particle and $m_a$, $m_b$ and $m_c$ are the masses of the
daughters;  $m_{ab}$,  $m_{ac}$, and $m_{bc}$
are the relevant invariant masses.
These expressions for angular distributions can be obtained from covariant-tensor
formalism or from orbital momentum partial waves decomposition using 
Legendre polynomials $P_L(\cos\theta)$,
where $\theta$ is the angle between particles $a$ and $c$ in the resonance $R$ rest frame.

For regular resonances such as
$K^*(892)$,
$\phi(1020)$,
$K^*(1410)$,
$K_2^*(1430)$, etc.,
we use the standard Breit-Wigner function,
\begin{equation}
\label{eqn:Breit-Wigner}
    {\cal  W}_R(m) = \frac{1}{m_R^2 - m^2 - i m_R \Gamma(m)}
\end{equation}
multiplied by the angular distribution,
$\Omega_{L}$, and the Blatt-Weisskopf form factors ${\cal  F}_D^L(q)$
and ${\cal  F}_R^L(q)$
for the $D$-meson and resonance $R$
decay vertexes, respectively.
We assume that 
the mass dependent width has the usual form
\begin{equation}
\label{eqn:Breit-Wigner_width}
 \Gamma(m) = \Gamma_R
                       \frac{m_R}{m}
                       \bigg(\frac{\rm P}{{\rm P}_R} \bigg)^{2L+1}
                       \big[{\cal F}_R^L({\rm P} \times r_R) \big]^2,
\end{equation}
where $\rm P$ is the decay products' momentum value in the decaying particle rest frame
and $r_R$ is the effective resonance radius.
The form factors
${\cal  F}_D^L(q)$
and ${\cal  F}_R^L(q)$ in Eqs.~\ref{eqn:MatrixElement} and \ref{eqn:Breit-Wigner_width}
are defined in the Blatt-Weisskopf form \cite{Blatt-Weisskopf}
\begin{eqnarray}
   \label{eqn:Blatt-W_formfactors}
   L=0: & & {\cal F}_V^0(q) = 1,                             \label{eqn:formf_scalar}\\
   L=1: & & {\cal F}_V^1(q) = {\cal N}_V^1 \times \Big[1+q^2 \Big]^{-1/2},
                                                               \label{eqn:formf_vector}\\
   L=2: & & {\cal F}_V^2(q) = {\cal N}_V^2 \times \Big[9+3q^2+q^4\Big]^{-1/2},
                                                               \label{eqn:formf_tensor}
\end{eqnarray}
where the label $V$ stands for $D$ or $R$ decay vertex,
$q={\rm P}\times r_V$,
$r_V$ is an effective meson radius, and ${\cal N}_V^L$ is a normalization constant
defined by the condition ${\cal F}_V^L({\rm P}_R \times r_V) = 1$,
where ${\rm P}_R$ is the products' momentum value at $m=m_R$.

The ${\cal W}_R$ parameterization of the $f_0(980)$,
whose mass, $m_{f_0}$, is close to the $K\overline{K}$ production threshold,
uses the Flatt\'e \cite{Flatte} formula
\begin{equation}
\label{eqn:Flatte}
{\cal W}_R(m) = \frac{1}{m_R^2 - m^2 - i \sum_{ab} g^2_{Rab} \rho_{ab}(m)}
\end{equation}
where $ab$ stands for $\pi^0\pi^0$, $\pi^+\pi^-$, $K^+K^-$, and $K^0\overline{K^0}$,
and $\rho_{ab}(m) = 2{\rm P}_a/m$ is a phase space factor, calculated for the decay
products momentum, ${\rm P}_a$, in the resonance rest frame.
We use the following isospin relations for the coupling constants
$g_{f_0\pi^+\pi^-} = \sqrt{2/3} g_{f_0\pi\pi}$,  
$g_{f_0\pi^0\pi^0} = \sqrt{1/3} g_{f_0\pi\pi}$, and  
$g_{f_0K^0\overline{K^0}} = g_{f_0K^+K^-} = \sqrt{1/2} g_{f_0K\overline{K}}$. 
Their values, shown in Table~\ref{tab:expected_contributions},
are taken from the BES experiment~\cite{f0BES}.

We model a low mass $K^+\pi^-$ S wave, also known as $\kappa$ or $K(800)$, 
using a complex pole amplitude proposed in Ref.~\cite{Oller_2005},
\begin{equation}
\label{eqn:ComplexPole}
{\cal W}_{\kappa}(m) = \frac{1}{m^2_{\kappa} - m^2},
\end{equation}
where
$m_{\kappa} = (0.71 - i 0.32)$~GeV is a pole position in the complex 
$s=m^2(K^+\pi^-)$ plane
estimated from the results of several experiments.

In this analysis we use or test all known $K^-\pi^+$ and  $K^+K^-$
resonances recognized by the Particle Data Group (PDG)~\cite{PDG-2006}
which can be observed in the phase space of the $D_s^+ \to K^-K^+\pi^+$ decay.
These are listed in Table~\ref{tab:expected_contributions}.
\begin{table}[th]
\caption{\label{tab:expected_contributions} Parameters of contributing resonances.}
\begin{center}
\begin{tabular}{ l c c c }
\hline
\hline
Resonance      & $J^{PC}$ & Mass (MeV/$c^2$)   & Width (MeV/$c^2$) \\
\hline
\multicolumn{4}{|c|}{ $K\pi$ states }                              \\
\hline
$K^*(892)$     & $1^-$    & 896.00\PM0.25      & 50.3\PM0.6        \\
$K^*(1410)$    & $1^-$    & 1414\PM15          & 232\PM21          \\
$K_0^*(1430)$  & $0^+$    & 1414\PM6           & 290\PM21          \\
$K_2^*(1430)$  & $2^+$    & 1432.4\PM1.3       & 109\PM5           \\
$K^*(1680)$    & $1^-$    & 1717\PM27          & 322\PM110         \\
$\kappa$       & $0^+$    &$\Re\,m$=710         &$\Im\,m$=--310      \\
\hline	        
\multicolumn{4} {|c|} { $K^+K^-$ states }                          \\
\hline	        
$f_0(980)$     & $0^{++}$ & 965\PM10           & $g_{\pi\pi}$=406  \\
               &          &                    & $g_{KK}$=800      \\
$a_0(980)$     & $0^{++}$ & 999\PM1            & $g_{\eta\pi}$=620 \\
               &          &                    & $g_{KK}$=500      \\
$\phi(1020)$   & $1^{--}$ & 1019.460\PM0.019   & 4.26\PM0.05       \\
$f_2(1270)$    & $2^{++}$ & 1275.4\PM1.1       & 185.2$^{+3.1}_{-2.5}$ \\
$a_2(1320)$    & $2^{++}$ & 1318.3\PM0.6       & 107\PM5           \\
$f_0(1370)$    & $0^{++}$ & 1200 to 1500       & 200 to 500        \\
$a_0(1450)$    & $0^{++}$ & 1474\PM19          & 265\PM13          \\
$f_0(1500)$    & $0^{++}$ & 1507\PM5           & 109\PM7           \\
$f_2(1525)$    & $2^{++}$ & 1525\PM5           & 73$^{+6}_{-5}$    \\
$f_0(1710)$    & $0^{++}$ & 1718\PM6           & 137\PM8           \\
$\phi(1680)$   & $1^{--}$ & 1680\PM20          & 150\PM50          \\ \hline \hline
\end{tabular}
\end{center}
\end{table}
One could expect a contribution in the $K^+K^-$ mass spectrum from
the $f_0(980)$ and $a_0(980)$ scalar resonances.
Their $K^+K^-$ mass spectra
have similar, but not well defined shapes. 
If both amplitudes are allowed to float simultaneously in the fit,
they show a huge destructive interference, which is sensitive
to their shape parameters.
The $f_0(980)$ contribution dominates \cite{PDG-2006} in 
the $D_s^+ \to \pi^+\pi^+\pi^-$ decay,
which has a large branching fraction, ${\cal B}(D_s^+ \to \pi^+\pi^+\pi^-)=(1.22 \pm 0.23)\%$.
The relevant coupled channel of the $a_0(980)$ has not been observed in
the $D_s^+ \to \eta \pi^0 \pi^+$ decay.
In this analysis we consider the $f_0(980)$ contribution only.

\section{Fits to Data}
\label{sec:fits_to_data}

First, we analyze our data with the model used by E687~\cite{E687}.
Their isobar model contains five contributions, 
$K^*(892)^0 K^+$, 
$\phi(1020)\pi^+$, 
$K_0^*(1430)K^+$, 
$f_0(980)\pi^+$, and 
$f_0(1710)\pi^+$.
In our analysis of $D^+\to K^-\pi^+\pi^+$ and $D^+\to K^-K^+\pi^+$ 
decays we find a $K^*(892)$ width that is smaller than the world average
value from the PDG~\cite{PDG-2006}.  Thus we let the mass
and width of $K^*(892)$ float in the fit.
Results are shown in
Table~\ref{tab:comp_E687}.  In this table and all succeeding tables, 
the units of the amplitudes are arbitrary (a.u.).
We find that the sign of the $\phi(1020)$ contribution is opposite to the sign obtained by E687,
but all other results are consistent within quoted uncertainties.
We find that this fit to our data sample has a poor $\chi^2/\nu$, where $\nu$ is the number
of degrees of freedom, giving a very small fit probability.
The $\chi^2$ is calculated over adaptive bins,  
similar to our previous analysis~\cite{DptoKpipi}.
This model does not represent our data well especially in the range of
$1.1<m^2_{KK}<1.5$~GeV$^2/c^4$.
\begin{table}[th]
\caption{\label{tab:comp_E687}
Comparison of CLEO-c results with E687 using the E687 isobar model.
Shown are the fitted magnitudes, $a$ in arbitrary units, the phases ($\phi$) in degrees,
defined relative to the $K^*(892)^0 \pi^+$ amplitude, and the fit fractions (FF).}
\begin{center}
\begin{tabular}{ l lcc }
\hline
\hline
Mode & Parameter        & E687               & CLEO-c  [PDG]        \\
\hline                                        
${\overline K}^*(892)^0 K^+$
     & $a$              & (fixed)            & 1 (fixed)            \\ %
     & $\phi$ $(^\circ)$& 0 (fixed)          & 0 (fixed)            \\ %
     & m (MeV/$c^2$)    &                    & 895.8\PM0.5 [896.00\PM0.25]  \\ %
     &$\Gamma$ (MeV/$c^2$)&                  & 44.2\PM1.0  [50.3\PM0.06]    \\ %
     & FF   (\%)        & 47.8\PM4.6\PM4.0   & 48.2\PM1.2           \\ %
\hline
${\overline K}_0^*(1430)K^+$
     & $a$              & N/A                & 1.76\PM0.12          \\ %
     & $\phi$ $(^\circ)$& 152\PM40\PM39      & 145\PM8              \\ %
     & FF   (\%)        & 9.3\PM3.2\PM3.2    & 5.3\PM0.7            \\ %
\hline
$\phi(1020)\pi^+$
     & $a$              & N/A                & 1.15\PM0.02          \\ %
     & $\phi$ $(^\circ)$& 178\PM20\PM24      & --15\PM4             \\ %
     & FF   (\%)        & 39.6\PM3.3\PM4.7   & 42.7\PM1.3           \\ %
\hline
$f_0(980)\pi^+$
     & $a$              & N/A                & 3.67\PM0.13          \\ %
     & $\phi$ $(^\circ)$& 159\PM22\PM16      & 156\PM3              \\ %
     & FF   (\%)        & 11.0\PM3.5\PM2.6   & 16.8\PM1.1           \\ %
\hline
$f_0(1710)\pi^+$
     & $a$              & N/A                & 1.27\PM0.07          \\ %
     & $\phi$ $(^\circ)$& 110\PM20\PM17      & 102\PM4              \\ %
     & FF   (\%)        & 3.4\PM2.3\PM3.5    & 4.4\PM0.4            \\ %
\hline
$\sum$ FF (\%) &        & 111.1              & 117.3\PM2.2          \\ %
Number of events on DP   &  &                    & 14400                \\ %
Number of Signal events  &  & 701\PM36           & 12226\PM22           \\ %
\hline
Goodness
     & $\chi^2 / \nu$   & 50.2/33            & 278/119              \\ %
\hline
\hline
\end{tabular}
\end{center}
\end{table}


The E687 model contains five resonances. 
Two of them, $K^*(892)$ and $\phi(1020)$,  are clearly seen on the Dalitz plot. 
The other three, $K_0^*(1430)$, $f_0(980)$, and $f_0(1710)$,
are too wide to be easily discerned.
To check their significance we remove them one-by-one
from the total amplitude and check the fit results.
In all fits where we remove one resonance
the fit quality is degraded, increasing $\chi^2 / \nu$ by more than $0.6$,
compared to our central fit.
Thus, we assume that all five resonances from E687 model
are significant.

\newcommand{\mKst} {$m_{K^*(892)}$}      
\newcommand{\wKst} {$\Gamma_{K^*(892)}$} 
\newcommand{\AKstZ}{$a_{K_0^*(1430)}$ (a.u.)}
\newcommand{\PKstZ}{$\phi_{K_0^*(1430)}$ ($^\circ$)}
\newcommand{\Afo}  {$a_{f_0(980)}$ (a.u.)}  
\newcommand{\Pfo}  {$\phi_{f_0(980)}$ ($^\circ$)}  
\newcommand{\Aphi} {$a_{\phi(1020)}$ (a.u.)} 
\newcommand{\Pphi} {$\phi_{\phi(1020)}$ ($^\circ$)} 
\newcommand{\AfoH} {$a_{f_0(1710)}$ (a.u.)} 
\newcommand{\PfoH} {$\phi_{f_0(1710)}$ ($^\circ$)}   
\newcommand{\AfoA} {$a_{f_0(1370)}$ (a.u.)} 
\newcommand{\PfoA} {$\phi_{f_0(1370)}$ ($^\circ$)}   
\newcommand{\Aadd} {$a_{\rm add}$ (a.u.)} 
\newcommand{\Padd} {$\phi_{\rm add}$ ($^\circ$)}
\newcommand{\ChSq} {$\chi ^2 / \nu$}
\newcommand{\Pear} {Pearson / $\nu$}
\newcommand{\Pois} {Poisson / $\nu$}
\newcommand{\Prob} {Prob.(\%)}
\newcommand{\EvDP} {Events on DP} 
\newcommand{\FadA} {FF$_{\rm add}$ (\%)}    
\newcommand{\Fadd} {FF$_{\rm add}$ (\%) @ 90\% C.L.}    
\newcommand{\FKst} {FF[$K^*(892)$] (\%)}    
\newcommand{\FKstZ}{FF[$K_0^*(1430)$] (\%)}    
\newcommand{\Fphi} {FF[$\phi(1020)$] (\%)}    
\newcommand{\Ffo}  {FF[$f_0(980)$] (\%)}    
\newcommand{\FfoH} {FF[$f_0(1710)$] (\%)}    
\newcommand{\FfoA} {FF[$f_0(1370)$] (\%)}    
\newcommand{\SumF} {$\sum_R$ FF$_R$ (\%)}    
\newcommand{\fsig} {$f_{\rm sig}$}

In order to get better consistency between the model and data, 
we try to improve the E687 model by adding contributions
from the other known resonances listed in 
Table~\ref{tab:expected_contributions}.
The results of these fits are shown in
Tables~\ref{tab:add_Kpi_resonance} and~\ref{tab:add_KK_resonance}
as a variation of the fit parameters with respect to the central case.
In all cases the fit quality is improved and each additional 
resonance has a significant magnitude.
We conclude that the five resonance model based on E687 results does
not fully describe the data sample. 
The largest fit quality improvement is achieved in the case of additional 
S-wave contributions: $f_0(1370)$; non-resonant ($NR$); $a_0(1450)$; and $\kappa$. 
Adding the $f_0(1370)$ contribution gives the largest improvement of the
fit quality, $\Delta\chi^2 = -100$ for two fewer degrees of freedom.

\begin{table}[th]
\caption{\label{tab:add_Kpi_resonance} Fits to CLEO-c data using the E687 model
                                       with additional $K^-\pi^+$ resonances.
                                       For the contributions that do not change
                                       the entries in the table are changes from the E687 model.}
\begin{center}
\begin{tabular}{lcccccc}
\hline
\hline
Parameter& E687 Model  & $NR$    &$K^*(1410)$&$K_2^*(1430)$&$K^*(1680)$ &$\kappa$   \\  
\hline		       	      		               
  \mKst  & 895.8\PM0.5 &   0.0   & --0.4     &--0.1        &--1.2       &--0.9      \\
  \wKst  &  44.2\PM1.0 &   0.4   & --1.3     &  0.3        &--2.1       &--0.3      \\
  \AKstZ &  1.76\PM0.12& --1.16  & --0.02    &  0.14       &  0.05      &--0.58     \\
  \PKstZ &   145\PM8   & --4.2   &4          &  7.3        &--4         &--7        \\
  \Afo   &  3.67\PM0.13&   1.64  &  0.28     &--0.19       &  0.69      &  0.91     \\
  \Pfo   &   156\PM3   &  41     & --2.2     &  4.3        &--0.78      & 29        \\
  \Aphi  &  1.15\PM0.02& --0.02  &  0.04     &  0.003      &  0.06      &--0.01     \\
  \Pphi  &  --15\PM4   &  32     & --13      &  0.6        &--10.4      & 26        \\
  \AfoH  &  1.27\PM0.07& --0.83  &  0.06     &--0.07       &  0.22      &--0.87     \\
  \PfoH  &   102\PM4   & --27    & --9.4     &  3.0        &--6.7       &--15       \\
\hline		       	      		        
   \Aadd &             &5.2\PM0.4&1.77\PM0.21&0.92\PM0.15  &6.3\PM0.9   &2.27\PM0.17\\
   \Padd &             &193\PM4  &  93\PM6   &-179\PM16    &117\PM9     &51\PM4     \\
\hline		       	      		        
   \ChSq &278/119      & 192/117 & 249/117   & 241/117     & 256/117    & 200/117   \\
\hline
\hline
\end{tabular}
\end{center}
\end{table}
\begin{table}[th]
\caption{\label{tab:add_KK_resonance} Fits to CLEO-c data using the E687 model with additional $K^+K^-$ resonances.
                                       For the contributions that do not change
                                       the entries in the table are changes from the E687 model.}
\begin{center}
\begin{tabular}{lcccccccc}
\hline
\hline
Parameter& E687 Model &$f_2(1270)$&$a_2(1320)$&$f_0(1370)$&$f_0(1500)$&$f_2(1525)$&$a_0(1450)$&$\phi(1680)$\\
\hline		         		      		               	                                   
  \mKst & 895.8\PM0.5 &--0.4      &--0.1      &--0.9      &--0.5      &  0.0      & --0.8     & 0.1        \\
  \wKst &  44.2\PM1.0 &  2.3      &  2.4      &  1.5      &  0.6      &  0.6      &   1.0     & 1.2        \\
  \AKstZ&  1.76\PM0.12&  0.11     &  0.08     &--0.25     &--0.03     &--0.16     & --0.22    &--0.18      \\
  \PKstZ&   145\PM8   &--32       &--28       &  1.0      &--15       &  1.7      & --15      &18          \\
  \Afo  &  3.67\PM0.13&  0.29     &  0.26     &  1.05     &  0.52     &  0.03     &   1.09    & 0.20       \\
  \Pfo  &   156\PM3   &--2        &--1.6      &  1.3      &  2.3      &  0.22     &   3.8     & 10.5       \\
  \Aphi &  1.15\PM0.02&--0.03     &--0.04     &--0.02     &--0.003    &--0.02     & --0.007   &--0.012     \\
  \Pphi &  --15\PM4   &--7        &--6.3      &  7.2      &--0.6      &  1.5      &   4.3     &13.2        \\
  \AfoH &  1.27\PM0.07&  0.08     &  0.07     &--0.16     &  0.17     &--0.04     &   0.03    &--0.018     \\
  \PfoH &   102\PM4   &  7        &  4.7      &--13       &--4.1      &--3.8      &--17       & 5.3        \\
\hline			  	      		      	                                           
   \Aadd&             &0.64\PM0.09&0.45\PM0.06&1.15\PM0.09&0.50\PM0.05&0.50\PM0.07&1.32\PM0.10&1.04\PM0.17 \\
   \Padd&             &17\PM9     &  40\PM8   &53\PM5     &132\PM7    &173\PM10   &103\PM5    &--4\PM11    \\
\hline				  	      		      	                                           
   \ChSq&278/119      & 237/117   & 237/117   & 178/117   & 229/117   & 249/117   & 192/117   & 256/117    \\
\hline
\hline
\end{tabular}
\end{center}
\end{table}

\begin{figure}[th]
  \includegraphics[width=72mm]{\FIGDIR/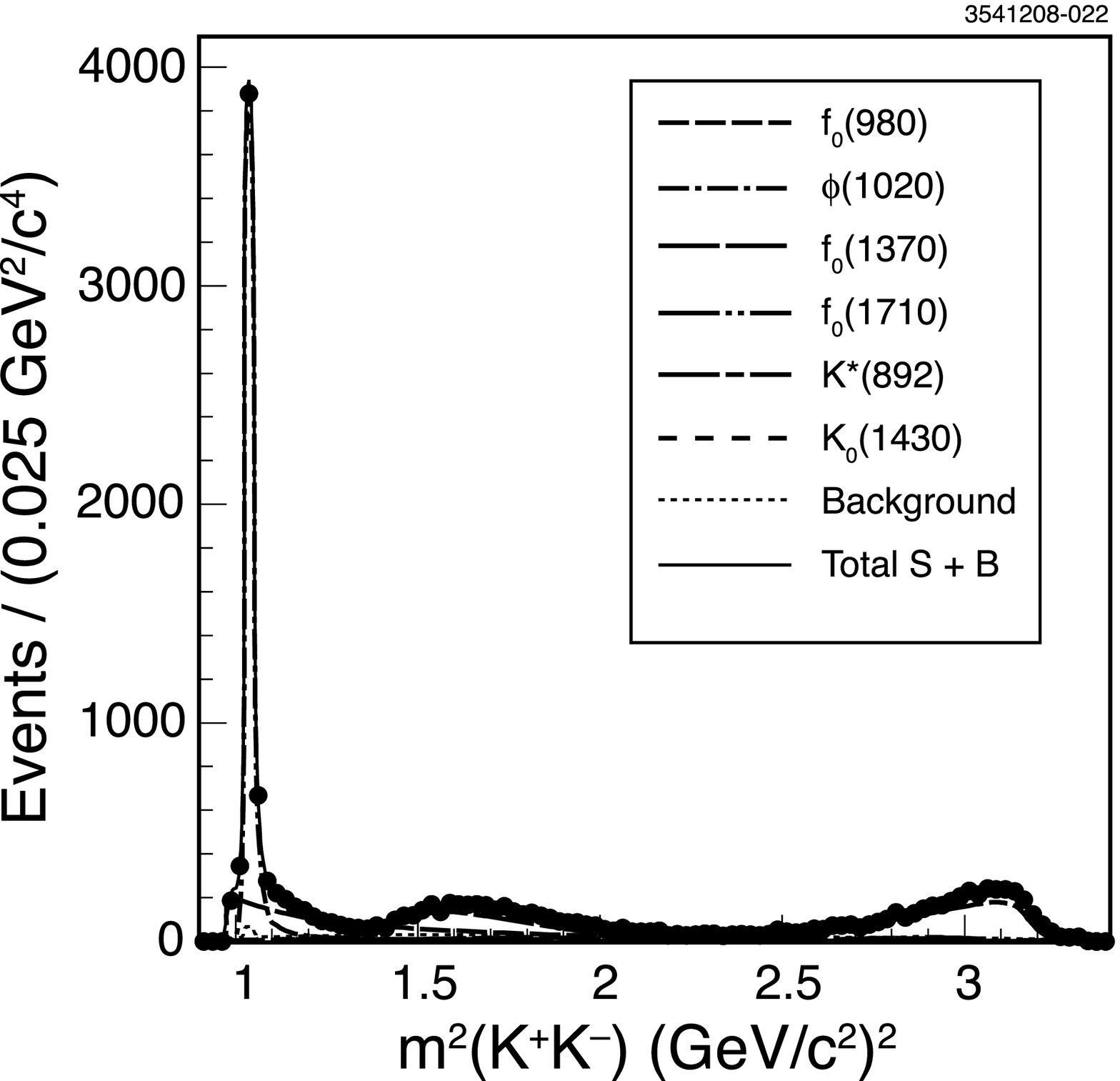} \hfill
  \includegraphics[width=72mm]{\FIGDIR/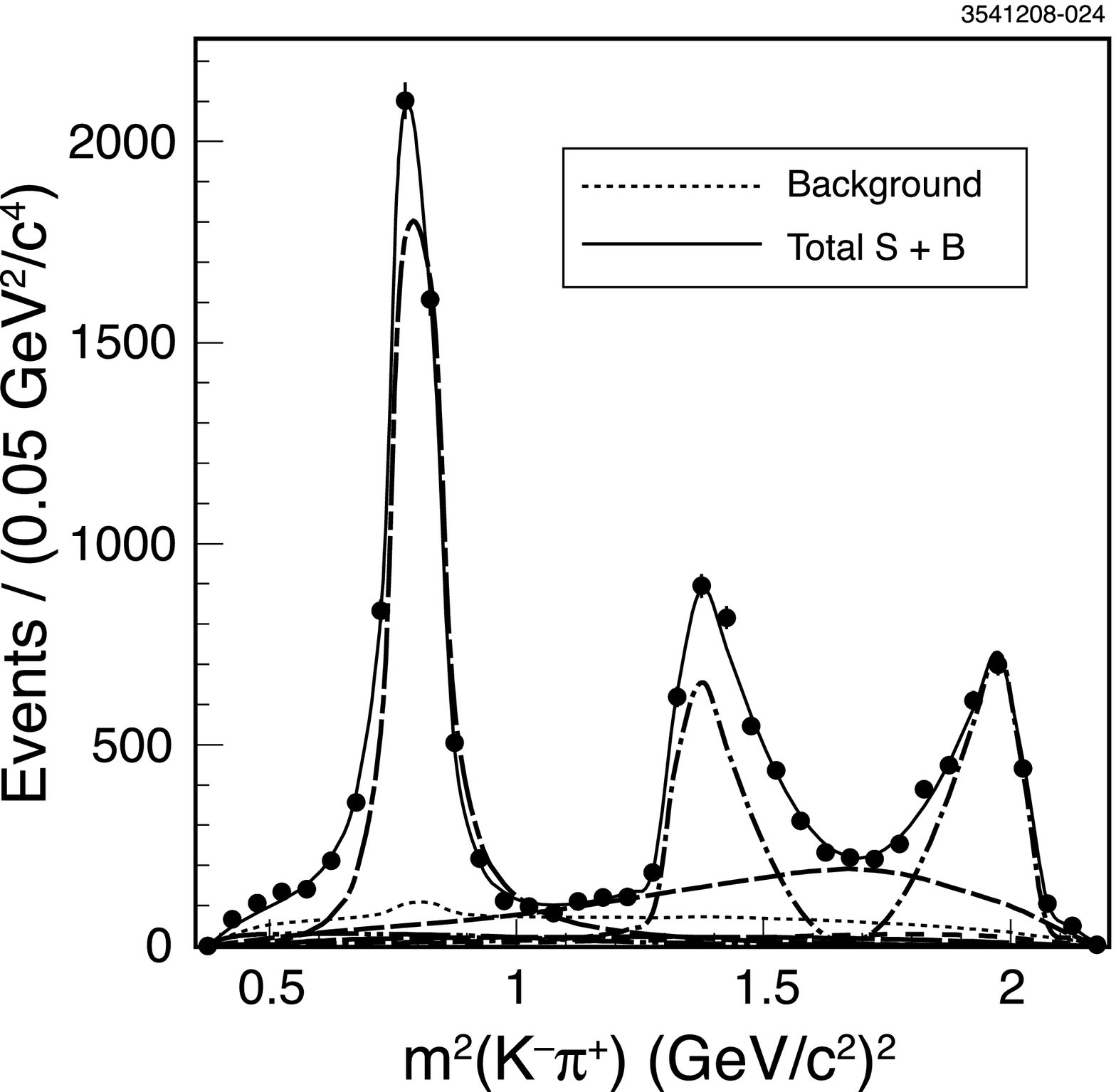} 
  \includegraphics[width=72mm]{\FIGDIR/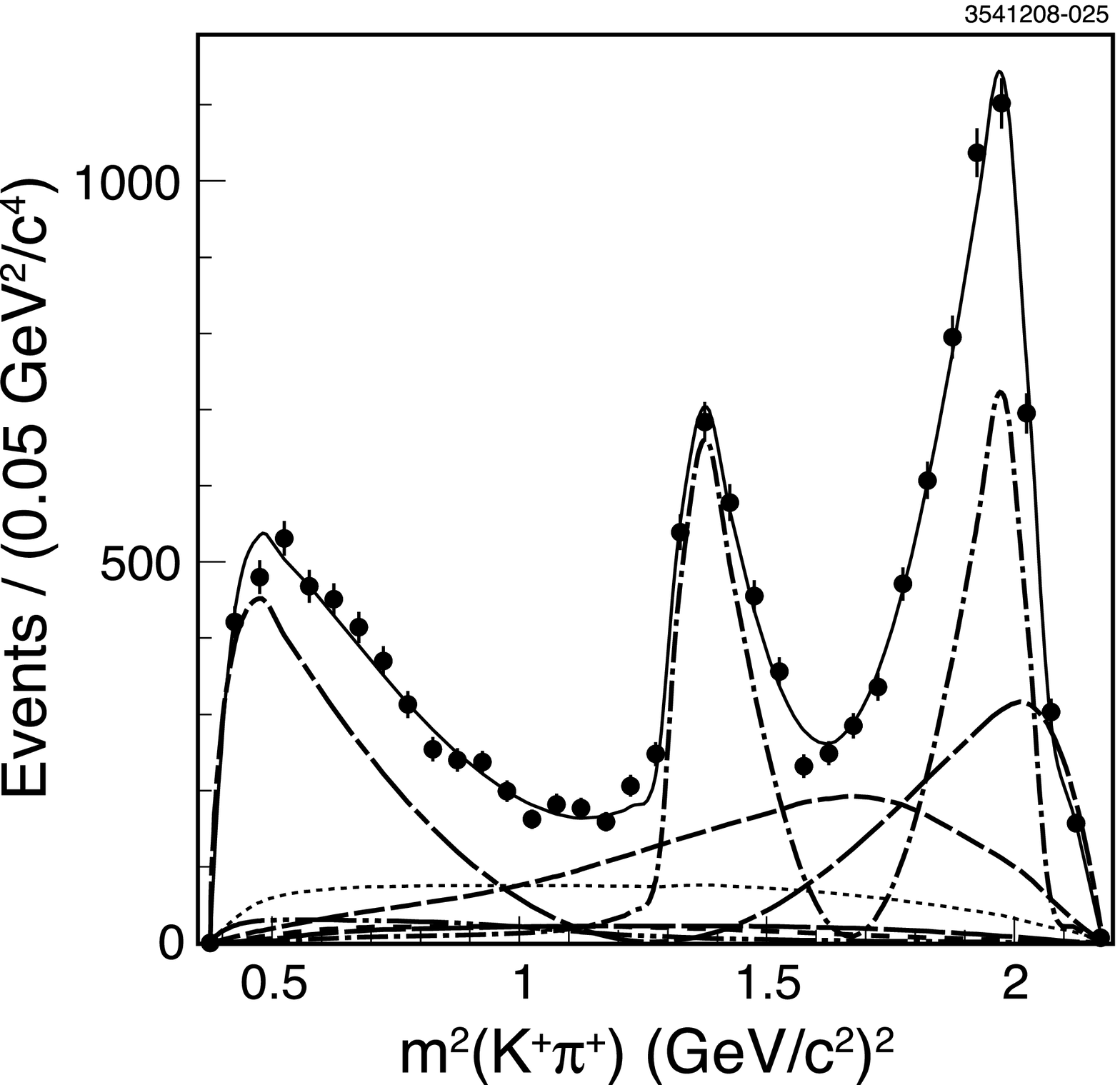} \hfill
  \includegraphics[width=72mm]{\FIGDIR/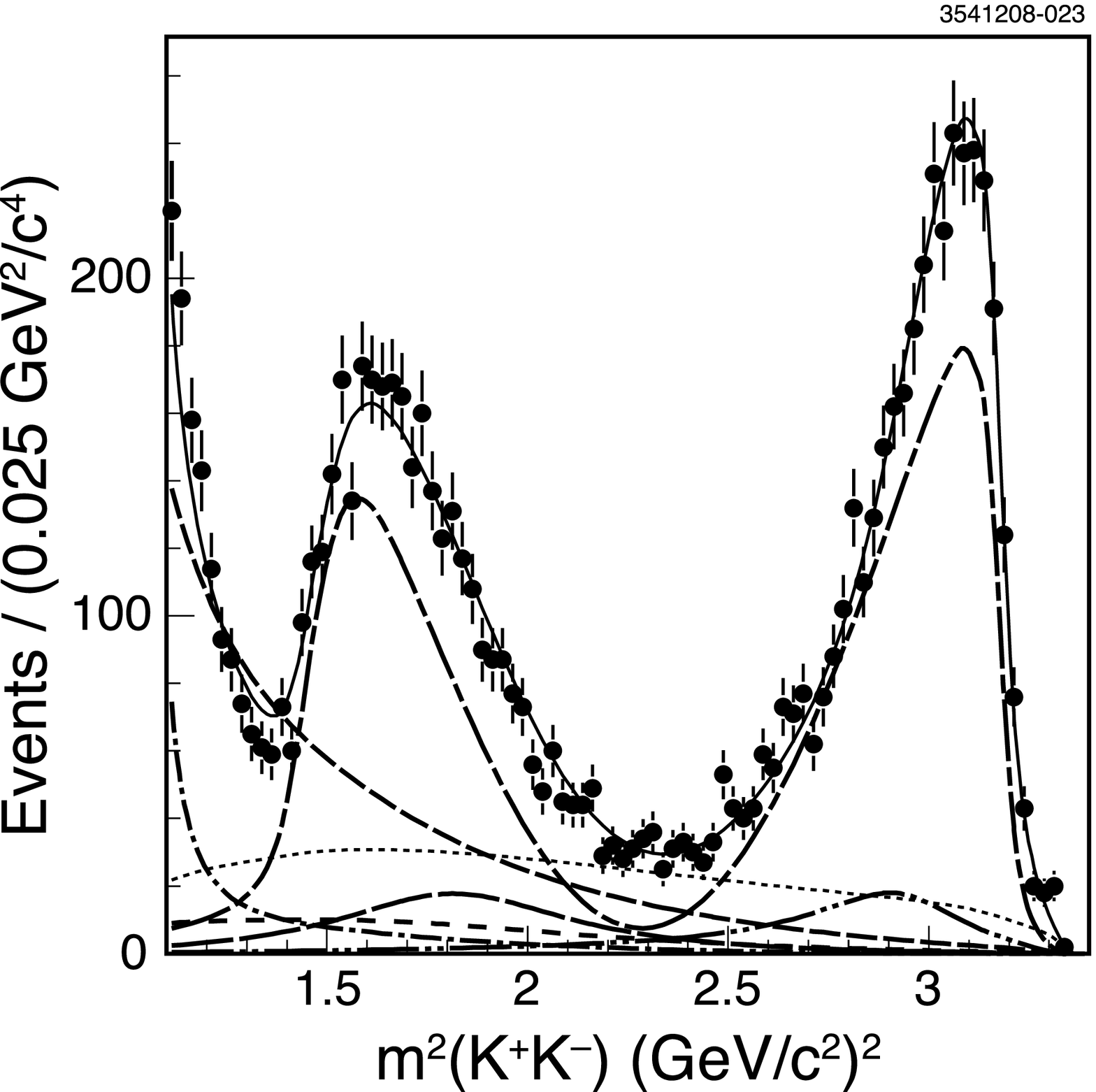} 
  \caption{\label{fig:data_proj_ModelA} Fit to data for Model A, and projections of the Dalitz plot.
                                        The final plot shows the $m^2(KK)$ projection of
                                        Dalitz plot for values of $m^2(KK)$ larger than the
                                        contribution from the $\phi(1020)$.} 
\end{figure}

We consider a six-resonance model, called Model A, containing 
$K^*(892)^0 K^+$, 
$\phi(1020)\pi^+$, 
$K_0^*(1430)K^+$, 
$f_0(980)\pi^+$, 
$f_0(1710)\pi^+$,
and 
$f_0(1370)\pi^+$ contributions. 
Model A is simply the E687 isobar model with an additional $f_0(1370)\pi^+$ 
contribution.
Results with this model and fit projections are shown in Fig.~\ref{fig:data_proj_ModelA}.
We repeat the previous procedure and
include one-by-one additional resonance
and check the significance of its parameters and consistency of the p.d.f. with our data sample. 
Results are shown in 
Tables~\ref{tab:add_Kpi_to_ModelA} and~\ref{tab:add_KK_to_ModelA}.
For Model A we do not find any additional resonances with significant magnitude,
the fit quality does not significantly improve, and thus we take this model
for our central result.
For each additional resonance we estimate an upper limit on its fit fraction
at 90\% confidence level, as also shown in 
Tables~\ref{tab:add_Kpi_to_ModelA} and \ref{tab:add_KK_to_ModelA}.
We conclude that the six-resonance Model A p.d.f. gives a good description of our data sample. 
\begin{table}[th]
\caption{\label{tab:add_Kpi_to_ModelA} Fits to data using Model A with additional non-resonant or 
                                       $K^+\pi^-$ resonance.
                                       For the contributions that do not change
                                       the entries in the table are changes from Model A.}
\begin{center}
\begin{tabular}{lcccccc}
\hline
\hline
Parameter& Model A     & $NR$    &$K^*(1410)$&$K_2^*(1430)$&$K^*(1680)$ &$\kappa$   \\  
\hline		         		      		
  \mKst  & 894.9\PM0.5 & 0.3     & 0.1       & 0.2         & 0.2        &-0.1       \\
  \wKst  &  45.7\PM1.1 &-0.1     & 0.1       & 0.8         & 0.6        &-0.3       \\
  \AKstZ &  1.51\PM0.11&-0.1878  &-0.0245    & 0.0603      &-0.1434     & 0.2685    \\
  \PKstZ &   146\PM8   &-10.833  & 0.4446    &-4.8755      & 2.3676     &-7.6608    \\
  \Afo   &  4.72\PM0.18&-0.0529  & 0.0057    & 0.2566      &-0.2530     &-0.2078    \\
  \Pfo   &   157\PM3   & 8.1153  &-0.7457    & 1.0875      & 1.7545     &-4.5506    \\
  \Aphi  &  1.13\PM0.02&-0.0005  &-0.0001    &-0.0096      &-0.0159     & 0.0047    \\
  \Pphi  &   --8\PM4   & 3.9973  &-0.1144    &-4.8349      & 5.2172     &-5.0235    \\
  \AfoA  &  1.15\PM0.09&-0.0979  &-0.0055    & 0.0535      & 0.0103     & 0.0890    \\
  \PfoA  &    53\PM5   & 5.5500  &-1.6829    &-4.4427      & 3.2688     &-11.386    \\
  \AfoH  &  1.11\PM0.07&-0.1502  &-0.0093    &-0.0157      &-0.0442     &-0.0940    \\
  \PfoH  &    89\PM5   &-7.3126  &-1.2087    & 3.7678      & 2.4526     &-6.2195    \\
\hline				  	      		        
   \Aadd &     0       &1.3\PM0.6&0.10\PM0.13&1.00\PM0.26  &2.18\PM1.33 &0.50\PM0.18\\
   \Padd &     0       &-147\PM19&-3\PM119   &105\PM11     &--72\PM13   &163\PM25   \\
   \FadA &     0       &1.5\PM1.4&0.01\PM0.03&0.40\PM0.22  &0.30\PM0.44 &0.40\PM0.32\\
   \Fadd &     0       & $<$3.3\%& $<$0.05\% & $<$0.7\%    & $<$0.9\%   & $<$0.8\%  \\
\hline				  	      		        
   \FKst & 47.4\PM1.5  & 47.5    & 47.5      & 47.8        & 48.3       & 47.5      \\
   \FKstZ&  3.9\PM0.5  &  3.0    &  3.8      &  4.4        &  3.3       &  5.5      \\
   \Ffo  & 28.2\PM1.9  & 27.7    & 28.4      & 32.3        & 26.2       & 25.7      \\
   \Fphi & 42.2\PM1.6  & 41.9    & 42.1      & 42.3        & 42.1       & 42.1      \\
   \FfoA &  4.3\PM0.6  &  3.5    &  4.2      &  4.8        &  4.5       &  4.9      \\
   \FfoH &  3.4\PM0.5  &  2.6    &  3.4      &  3.4        &  3.3       &  2.9      \\
   \SumF & 129.5       & 127.8   & 129.4     & 135.4       & 127.9      & 129.0     \\
\hline				  	      		        
   \ChSq & 178/117     & 174/115 & 177/115   & 170/115     & 175/115    & 173/115   \\
\hline
\hline
\end{tabular}
\end{center}
\end{table}
\begin{table}[th]
\caption{\label{tab:add_KK_to_ModelA} Fits using Model A with additional $K^+K^-$ resonance.
                                       For the contributions that do not change
                                       the entries in the table are changes from Model A.}
\begin{center}
\begin{tabular}{lccccccc}
\hline
\hline
Parameter& Model A    &$f_2(1270)$&$a_2(1320)$&$f_0(1500)$&$f_2(1525)$&$a_0(1450)$&$\phi(1680)$ \\
\hline		         		      		               	                         
  \mKst & 894.9\PM0.5 &-0.5       &-0.3       &-0.1       &-0.2       &-0.1       &-0.1         \\
  \wKst &  45.7\PM1.1 & 1.2       & 1.2       &-0.1       & 0.2       &-0.1       &-0.2         \\
  \AKstZ&  1.51\PM0.11&-0.0518    &-0.0587    &-0.0060    &-0.0822    &-0.0210    &-0.0152      \\
  \PKstZ&   146\PM8   &-13.610    &-7.5258    & 1.1483    & 0.1662    & 2.4740    &-0.8833      \\
  \Afo  &  4.72\PM0.18& 0.0864    &-0.0037    & 0.0521    &-0.0239    & 0.1123    & 0.0113      \\
  \Pfo  &   157\PM3   &-0.6746    &-0.6856    & 0.6617    &-0.3009    & 1.1151    &-0.1360      \\
  \Aphi &  1.13\PM0.02&-0.0105    &-0.0126    & 0.0058    &-0.0058    & 0.0068    & 0.0056      \\
  \Pphi &   --8\PM4   &-2.1292    &-1.5385    & 0.5046    &-0.1244    & 1.2202    &-0.4788      \\
  \AfoA &  1.15\PM0.09&-0.0176    &-0.0343    & 0.0336    &-0.0168    & 0.0150    &-0.0039      \\
  \PfoA &    53\PM5   & 1.0892    &-0.3964    & 3.8125    & 1.4021    &14.6004    & 0.3390      \\
  \AfoH &  1.11\PM0.07& 0.0041    &-0.0165    &-0.0161    &-0.0100    &-0.0533    & 0.0007      \\
  \PfoH &    89\PM5   & 4.7785    & 2.7846    &-1.9584    &-2.2626    &-3.6665    &-0.9276      \\
\hline	               	  	      		      	                                           
   \Aadd&     0       &0.40\PM0.09&0.26\PM0.06&0.07\PM0.04&0.23\PM0.08&0.37\PM0.28&0.10\PM0.16  \\
   \Padd&     0       &22\PM14    &  51\PM15  &37\PM66    &180\PM26   &24\PM17    &--93\PM122   \\
   \FadA&     0       &0.24\PM0.11&0.20\PM0.09&0.04\PM0.10&0.09\PM0.05&0.38\PM0.60&0.008\PM0.031\\
   \Fadd&     0       & $<$0.4\%  & $<$0.3\%  & $<$0.17\% & $<$0.16\% & $<$1.2\%  & $<$0.05\%   \\
\hline				  	      		        
   \FKst & 47.4\PM1.5 & 47.2      & 47.4      & 47.3      & 48.0      & 47.3      & 47.4        \\
   \FKstZ&  3.9\PM0.5 &  3.8      &  3.7      &  3.9      &  3.6      &  3.8      &  3.8        \\
   \Ffo  & 28.2\PM1.9 & 30.0      & 29.0      & 28.8      & 28.4      & 29.4      & 28.2        \\
   \Fphi & 42.2\PM1.6 & 42.1      & 42.2      & 42.2      & 42.1      & 42.2      & 42.1        \\
   \FfoA &  4.3\PM0.6 &  4.2      &  4.1      &  4.5      &  4.2      &  4.3      &  4.2        \\
   \FfoH &  3.4\PM0.5 &  3.5      &  3.4      &  3.3      &  3.4      &  3.1      &  3.4        \\
   \SumF & 129.5      & 131.1     & 130.2     & 130.0     & 129.8     & 130.5     & 129.3       \\
\hline	                  		  	      		      	                               
   \ChSq&  178/117    & 169/115   & 170/115   & 177/115   & 172/115   & 176/115   & 178/115     \\
\hline
\hline
\end{tabular}
\end{center}
\end{table}

For Model A we test the resonance shape parameters by
floating the mass and width, or two coupling constants in case of $f_0(980)$,
for each resonance. 
Results of these fits are shown in
Tables~\ref{tab:resonance_parameters} and
      ~\ref{tab:fits_for_resonance_parameters}.
We find that all parameters are consistent with their central fit values used in the fit with Model A.
\begin{table}[th]
\caption{\label{tab:resonance_parameters} Optimal resonance parameters.  The uncertainties
                                          for the CLEO-c results are statistical only.}
\begin{center}
\begin{tabular}{lcccc}
\hline
\hline
Resonance      & Parameter (MeV/$c^2$)& Central Fit      & Floated       &PDG~\cite{PDG-2006}\\ \hline
	   
$K^*(892)$     & $m$                  & 895.8\PM0.5      & 895.8\PM0.5   & 896.00\PM0.25     \\      
               & $\Gamma$             &  44.2\PM1.0      &  44.2\PM1.0   & 50.3\PM0.6        \\
	   								                    
$K_0^*(1430)$  & $m$                  & 1414             & 1422\PM23     & 1414\PM6          \\     
               & $\Gamma$             & 290              &  239\PM48     & 290\PM21          \\
	   								                    
$f_0(980)$     & $m$                  & 965              & 933\PM21      & 980\PM10          \\    
               & $g_{\pi\pi}$         & 406              & 393\PM36      &$\Gamma$=40 to 100 \\
               & $g_{KK}$             & 800              & 557\PM88      &                   \\
	   								                      
$\phi(1020)$   & $m$                  &1019.460          & 1019.64\PM0.05& 1019.460\PM0.019  \\  
               & $\Gamma$             & 4.26             & 4.780\PM0.14   & 4.26\PM0.05       \\
	   								                    
$f_0(1370)$    & $m$                  &1350              & 1315\PM34     & 1200 to 1500      \\
               & $\Gamma$             & 265              &  276\PM39     & 200 to 500        \\

$f_0(1710)$    & $m$                  & 1718             & 1749\PM12     & 1718\PM6          \\
               & $\Gamma$             & 137              &  175\PM29     & 137\PM8           \\
\hline
\hline
\end{tabular}
\end{center}
\end{table}
\begin{table}[th]
\caption{\label{tab:fits_for_resonance_parameters} Fits to data using Model A 
                                                   with floating resonance parameters.
                                       After the first column of data
                                       the entries in the table are changes from Model A
                                       when the parameters of resonance at the top of the
                                       column are allowed to float.}
\begin{center}
\begin{tabular}{lcccccc}
\hline
\hline
Parameter& Model A     &$K^*(1430)$&$f_0(980)$  &$\phi(1020)$ &$f_0(1370)$ &$f_0(1710)$\\  
\hline		         		      		               
  \mKst  & 894.9\PM0.5 &-0.1       & 0          & 0.2         &-0.1        & 0.1       \\
  \wKst  &  45.7\PM1.1 &-0.1       & 0.2        & 0.1         & 0.0        &-0.5       \\
  \AKstZ &  1.51\PM0.11&-0.1449    &-0.1527     & 0.0256      & 0.0533     &-0.0305    \\
  \PKstZ &   146\PM8   & 8.6060    &-3.2558     &10.2102      & 7.5225     &-5.6685    \\
  \Afo   &  4.72\PM0.18&-0.0576    &-0.3873     &-0.3073      &-0.0540     & 0.1767    \\
  \Pfo   &   157\PM3   &-1.1202    &-13.584     & 0.0037      &-1.2207     & 3.4058    \\
  \Aphi  &  1.13\PM0.02& 0.0058    &-0.0018     & 0.0786      & 0.0037     & 0.0167    \\
  \Pphi  &   --8\PM4   &-0.8216    & 5.2291     & 1.5697      & 0.9613     & 1.3374    \\
  \AfoA  &  1.15\PM0.09& 0.0473    &-0.0319     &-0.0508      & 0.0293     &-0.1248    \\
  \PfoA  &    53\PM5   &-2.5387    & 4.8538     &-2.6304      &-17.247     & 3.0673    \\
  \AfoH  &  1.11\PM0.07&-0.0060    &-0.0096     &-0.0291      &-0.0656     & 0.4223    \\
  \PfoH  &    89\PM5   &-1.9306    &-1.2058     &-2.4148      & 0.0913     &20.0144    \\
\hline				  	      		        
   \FKst & 47.4\PM1.5  & 47.3      & 47.2       & 47.4        & 47.5       & 46.8      \\
   \FKstZ&  3.9\PM0.5  &  3.8      &  3.2       &  4.1        &  4.2       &  3.7      \\
   \Ffo  & 28.2\PM1.9  & 27.5      & 29.7       & 24.8        & 27.7       & 29.7      \\
   \Fphi & 42.2\PM1.6  & 42.2      & 41.8       & 43.3        & 42.2       & 42.0      \\
   \FfoA &  4.3\PM0.6  &  4.6      &  4.0       &  3.9        &  4.4       &  3.3      \\
   \FfoH &  3.4\PM0.5  &  3.4      &  3.4       &  3.3        &  3.0       &  4.1      \\
   \SumF & 129.5       & 128.8     & 129.2      & 126.8       & 129.0      & 129.5     \\
\hline	             		  	      		        
   \ChSq & 178/117     & 177/115   & 169/114    & 168/115     & 176/115    & 166/115   \\
\hline
\hline
\end{tabular}
\end{center}
\end{table}

To estimate systematic uncertainties of the fit parameters,
we apply numerous variations to the fitting procedure and look at the change
of the fit parameters from the central result.
We consider sub-samples where
the data is split into earlier and later datasets,
$D_s^+$ and $D_s^-$ decays, and selected using
tight and loose signal boxes.  These are
shown in Table~\ref{tab:systematic_A}. These results are obtained with
fixed parameters for efficiency and background functions 
from Tables~\ref{tab:Efficiency} and~\ref{tab:Background}.
We also consider fits with floating efficiency or
background parameters in Table~\ref{tab:systematic_A_eff_bkg_float}.
In these fits all polynomial coefficients for the efficiency or background
including resonance background amplitudes float freely, but we fit simultaneously 
two samples of events for data plus the signal MC efficiency or background box
to constrain the variation of the efficiency or background parameters. 
We also fit allowing the signal fraction to float, and
find $f_{\rm sig}=0.8495 \pm 0.0070$ which is consistent
with $0.8490$ used in the central fit.

We estimate a systematic uncertainty of the Model A fit parameters
by combining the fit results from 
Tables~\ref{tab:add_Kpi_to_ModelA},
\ref{tab:add_KK_to_ModelA}, 
\ref{tab:fits_for_resonance_parameters},
\ref{tab:systematic_A}, and
\ref{tab:systematic_A_eff_bkg_float}.
None of the systematic variations dominate the uncertainty.  The systematic
uncertainty is estimated as the mean change from the central fit
result, {\it $\delta$Mean},  added in quadrature to the RMS of all variations.  
The resulting systematic uncertainties on the parameters are given
in Table~\ref{tab:summary}.

\begin{table}[th]
\caption{\label{tab:systematic_A} Fits to a variety of data samples using Model A
                                  with central efficiency and background.
                                       After the first column of data
                                       the entries in the table are changes from Model A
                                       with the variation indicated at the top of the column.}
\scriptsize
\begin{center}
\begin{tabular}{lccccccccc}
\hline
\hline
Variation   & Central Fit & Early & Late & Only    & Only    & Tight    
& Loose
& Low Side & High Side       \\
Parameter   & Model A     & Data  & Data & $D_s^+$ & $D_s^-$ & $1\sigma \times 1\sigma$
& $3\sigma \times 3\sigma$
& Band & Band \\
\hline
  \mKst  & 894.9\PM0.5 &-0.4       & 3.0       &-0.7       & 0.7     &-0.2     & 0.2       &-1.2       &-1.4       \\
  \wKst  &  45.7\PM1.1 & 0.1       & 0.0       &-0.8       & 0.8     &-0.2     & 1.0       & 4.8       & 2.2       \\
  \AKstZ &  1.51\PM0.11& 0.0138    & 0.0177    &-0.0023    & 0.0398  &-0.0205  &-0.1276    &-0.8084    & 0.7309    \\
  \PKstZ &   146\PM8   &-10.971    & 9.7985    &-17.161    & 17.257  &-6.2148  & 14.408    & 18.400    &-66.057    \\
  \Afo   &  4.72\PM0.18& 0.3277    &-0.3513    & 0.0484    &-0.0416  &-0.0364  & 0.0244    & 0.6752    & 0.3610    \\
  \Pfo   &   157\PM3   &-1.3604    & 1.1808    &-6.3697    & 6.6295  &-4.5506  & 2.8515    & 3.1875    &-23.699    \\
  \Aphi  &  1.13\PM0.02& 0.0053    & 0.0008    & 0.0084    & 0.0011  & 0.0153  &-0.0049    & 0.0079    & 0.0210    \\
  \Pphi  &   --8\PM4   &-2.9134    & 2.2119    &-8.3156    & 8.5410  &-7.0696  & 5.5073    & 8.5766    &-35.140    \\
  \AfoA  &  1.15\PM0.09& 0.0976    &-0.1031    &-0.0131    & 0.0250  &-0.1193  & 0.1395    & 0.5111    & 0.3938    \\
  \PfoA  &    53\PM5   &-2.8318    & 2.2204    &-4.6088    & 2.4167  &-6.5716  &-1.3470    &-14.394    &-28.267    \\
  \AfoH  &  1.11\PM0.07& 0.0786    &-0.0830    &-0.0412    & 0.0483  & 0.0403  & 0.0070    & 0.1877    &-0.3847    \\
  \PfoH  &    89\PM5   &-3.3881    & 2.2247    & 0.1313    &-0.5966  & 0.7797  & 2.6467    & 16.146    &-5.0150    \\
\hline				  	            					                           
   \FKst & 47.4\PM1.5  & 47.2      & 47.7      & 47.9      & 46.7    & 47.2    & 46.8      & 43.4      & 48.0      \\
   \FKstZ&  3.9\PM0.5  &  4.0      &  4.1      &  3.9      &  4.2    &  3.8    &  3.3      &  0.9      &  8.3      \\
   \Ffo  & 28.2\PM1.9  & 32.1      & 24.4      & 28.6      & 27.9    & 27.6    & 28.8      & 37.6      & 31.5      \\
   \Fphi & 42.2\PM1.6  & 42.0      & 42.3      & 42.1      & 42.2    & 42.7    & 42.0      & 43.3      & 41.8      \\
   \FfoA &  4.3\PM0.6  &  5.0      &  3.5      &  4.1      &  4.5    &  3.4    &  5.4      &  9.0      &  7.4      \\
   \FfoH &  3.4\PM0.5  &  3.9      &  3.0      &  3.2      &  3.8    &  3.7    &  3.5      &  4.8      &  1.4      \\
   \SumF & 129.5       & 134.2     & 124.9     & 129.7     & 129.2   & 128.3   & 129.9     & 138.9     & 138.4     \\
\hline	             		  	            		                                                   
   \ChSq & 178/117     & 134/117   & 203/117   & 166/117   & 123/117 & 155/117 & 201/117   & 140/117   & 138/117   \\
   \EvDP & 14400       & 7334      & 7066      & 7233      & 7167    & 7200    & 19177     & 6682      & 7232      \\
   \fsig & 0.8490      & 0.8518    & 0.8466    & 0.8496    & 0.8497  & 0.9238  & 0.7484    & 0.4338    & 0.5696    \\
\hline    
\hline
\end{tabular}
\end{center}
\end{table}
\begin{table}[th]
\caption{\label{tab:systematic_A_eff_bkg_float} Fits to data using Model A
         with floating efficiency and background coefficients,
         fits with floating $f_{\rm sig}$, and with floating background
         coefficients $B_{K^*}$ and $B_{\phi}$ for the narrow resonance contributions
         to the background.                                       After the first column of data
                                       the entries in the table are changes from Model A
                                       with the variation indicated at the top of the column.}

\begin{center}
\begin{tabular}{lcccc}
\hline
\hline
Parameter& Model A     & Float $E_i$ & Float $B_i$ & Float $f_{\rm sig}$ \\
\hline		         		      	           
  \mKst  & 894.9\PM0.5 & 0         & 0.1       & 0         \\
  \wKst  &  45.7\PM1.1 & 0         &-0.2       & 0         \\
  \AKstZ &  1.51\PM0.11&-0.0018    &-0.0121    & 0.0023    \\
  \PKstZ &   146\PM8   & 0.1630    &-1.6971    & 0.2116    \\
  \Afo   &  4.72\PM0.18&-0.0026    &-0.0332    &-0.0043    \\
  \Pfo   &   157\PM3   & 0.3362    &-0.6851    & 0.2704    \\
  \Aphi  &  1.13\PM0.02& 0.0034    &-0.0007    & 0.0028    \\
  \Pphi  &   --8\PM4   & 0.1282    &-0.9907    &-0.0391    \\
  \AfoA  &  1.15\PM0.09&-0.0015    & 0.0112    & 0.0006    \\
  \PfoA  &    53\PM5   & 0.1323    &-0.5403    & 0.0792    \\
  \AfoH  &  1.11\PM0.07&-0.0007    &-0.0539    &-0.0038    \\
  \PfoH  &    89\PM5   &-0.2072    &-1.1088    &-0.3882    \\
\hline				  	      	           
   \FKst & 47.4\PM1.5  & 47.4      & 47.7      & 47.4      \\
   \FKstZ&  3.9\PM0.5  &  3.9      &  3.9      &  3.9      \\
   \Ffo  & 28.2\PM1.9  & 28.2      & 28.1      & 28.2      \\
   \Fphi & 42.2\PM1.6  & 42.2      & 42.2      & 42.2      \\
   \FfoA &  4.3\PM0.6  & 4.2       &  4.4      &  4.3      \\
   \FfoH &  3.4\PM0.5  & 3.4       &  3.1      &  3.4      \\
   \SumF & 129.5       & 129.4     & 129.3     & 129.4     \\
\hline               
   \ChSq & 178/117     & 679/562   & 270/188   & 178/116   \\
\hline    
\hline
\end{tabular}
\end{center}
\end{table}

\begin{table}[th]
\caption{\label{tab:summary} Summary of systematic cross checks for Model A.
                             Fit parameters are shown with their statistical 
                             and systematic uncertainty respectively.
			     The ``$\delta$Mean'' and ``RMS'' account for variation of the fit 
                             parameters in the systematic cross checks as discussed in the text. The ``Total'' is a quadratic
                             sum of ``$\delta$Mean'' and ``RMS'' and after rounding is the systematic uncertainty given
                             in the second column.
                             The results of the E687 Model are also shown for comparison.}

\begin{center}
\begin{tabular}{lccccc}
\hline
\hline
Parameter& Model A         &$\delta$Mean & RMS   & Total   & E687 Model  \\
\hline		         		      		                
  \mKst  & 894.9\PM0.5\PM0.7   &   0.088 & 0.654 & 0.660   & 895.8\PM0.5 \\
  \wKst  &  45.7\PM1.1\PM0.5   &   0.148 & 0.499 & 0.520   &  44.2\PM1.0 \\
  \AKstZ &  1.51\PM0.11\PM0.09 &  -0.024 & 0.089 & 0.092   &  1.76\PM0.12\\
  \PKstZ &   146\PM8\PM8       &  -0.623 & 8.442 & 8.465   &   145\PM8   \\
  \Afo   &  4.72\PM0.18\PM0.17 &  -0.029 & 0.167 & 0.170   &  3.67\PM0.13\\
  \Pfo   &   157\PM3\PM4       &  -0.343 & 4.036 & 4.051   &   156\PM3   \\
  \Aphi  &  1.13\PM0.02\PM0.02 &   0.004 & 0.017 & 0.018   &  1.15\PM0.02\\
  \Pphi  &   --8\PM4\PM4       &   0.081 & 3.850 & 3.851   &  --15\PM4   \\
  \AfoA  &  1.15\PM0.09\PM0.06 &  -0.003 & 0.063 & 0.063   &             \\
  \PfoA  &    53\PM5\PM6       &  -0.536 & 5.820 & 5.845   &             \\
  \AfoH  &  1.11\PM0.07\PM0.10 &  -0.004 & 0.098 & 0.098   &  1.27\PM0.07\\
  \PfoH  &    89\PM5\PM5       &   0.195 & 4.916 & 4.920   &   102\PM4   \\
\hline				                        
   \FKst & 47.4\PM1.5\PM0.4    &   0.016 & 0.357 & 0.4     & 48.2\PM1.2  \\
   \FKstZ&  3.9\PM0.5\PM0.5    &   0.036 & 0.460 & 0.5     &  5.3\PM0.7  \\
   \Ffo  & 28.2\PM1.9\PM1.8    &   0.096 & 1.792 & 1.8     & 16.8\PM1.1  \\
   \Fphi & 42.2\PM1.6\PM0.3    &   0.018 & 0.277 & 0.3     & 42.7\PM1.3  \\
   \FfoA &  4.3\PM0.6\PM0.5    &   0.044 & 0.488 & 0.5     &             \\
   \FfoH &  3.4\PM0.5\PM0.3    &   0.044 & 0.311 & 0.3     &  4.4\PM0.4  \\
   \SumF &129.5\PM4.4\PM2.0    &   0.020 & 1.981 & 2.0     &117.3\PM2.2  \\
\hline               
   \ChSq & 178/117             & \multicolumn{3}{c||}{}     &  278/119    \\
\hline    
\hline
\end{tabular}
\end{center}
\end{table}

\section{Conclusion}
\label{sec:summary}

We perform a Dalitz plot analysis of the $D_s^+\to K^+K^-\pi^+$ decay
with the CLEO-c data set of 586~pb$^{-1}$ of $e^+e^-$ collisions
accumulated at $\sqrt{s} = 4.17$~GeV.  This corresponds to
about 0.57 million $D_s^+D_s^{*-}$ pairs from which we select 14400
candidate events with a background of 15\%.
We compare our results with the previous measurement from E687
using the isobar model and find good agreement with the E687 parameters,
as shown in Table~\ref{tab:comp_E687}. 
We find that all resonances from E687 model are significant
and their exclusion degrades the fit quality.

However, the fit quality is signficantly improved if we
add an additional $K^+ K^-$ resonance to the model.
As shown in Tables~\ref{tab:add_Kpi_resonance}
and \ref{tab:add_KK_resonance}, almost any additional resonance or
non-resonant contribution improves the agreement with the data.
The best improvement is achieved if we add an $f_0(1370)\pi^+$
contribution. 
We find that a six-resonance model, containing contributions from
$K^*(892)^0 K^+$, 
$K_0^*(1430)K^+$, 
$f_0(980)\pi^+$, 
$\phi(1020)\pi^+$, 
$f_0(1370)\pi^+$,
and 
$f_0(1710)\pi^+$
resonances, gives better consistency with our data with $\chi^2/\nu = 178/117$.
Tables~\ref{tab:add_Kpi_to_ModelA} and \ref{tab:add_KK_to_ModelA}
show that any further additional resonance does not have a significant amplitude,
fit fraction, or significantly improve the fit quality and we give upper limits
on their fit fractions at the 90\% C.L.

In Table~\ref{tab:resonance_parameters} we show the resonance parameters
when they are allowed to float in the fit.  We find that the $K^*(892)$
width is 5~MeV/$c^2$ smaller than in PDG. This result is consistent 
with our observation in the $D^+ \to K^-\pi^+\pi^+$ analysis~\cite{DptoKpipi}.
Other resonance parameters are consistent with their values from the
PDG~\cite{PDG-2006} or the BES experiment~\cite{f0BES} for $f_0(980)$.

We estimate a systematic uncertainty on fit parameters from numerous 
fit variations, and 
Table~\ref{tab:summary} shows the final results on fit parameters
with their statistical and systematic uncertainties.

\section*{Acknowledgments}

We gratefully acknowledge the effort of the CESR staff
in providing us with excellent luminosity and running conditions.
D.~Cronin-Hennessy and A.~Ryd thank the A.P.~Sloan Foundation.
This work was supported by the National Science Foundation,
the U.S. Department of Energy,
the Natural Sciences and Engineering Research Council of Canada, and
the U.K. Science and Technology Facilities Council.

\end{document}